\documentclass[runningheads,fleqn]{cl2emult}

\usepackage{makeidx}  
\usepackage{graphicx} 
\usepackage{subeqnar} 
\usepackage{multicol} 
\usepackage{cropmark} 
\usepackage{phys}     
\makeindex            

\newcommand{\bu}{{\bf \hat u}}                                    %
\newcommand{\bR}{{\bf R}}                                         %
\newcommand{\bsigma}{\mbox{\boldmath $\sigma$}}                   %
\newcommand{\bTheta}{{\bf\Theta}}                                 %

\begin{document}

\title*{Ab initio theory of the interlayer exchange coupling}

\titlerunning{Interlayer Coupling}

\author{Josef Kudrnovsk\'y\inst{1,2}
\and V\'aclav Drchal\inst{1,2}
\and Ilja Turek\inst{3,2}
\and Patrick Bruno\inst{4}
\and Peter Dederichs\inst{5}
\and Peter Weinberger\inst{2}}
\authorrunning{Josef Kudrnovsk\'y et al.}

\institute{ Institute of Physics, Academy of Sciences of the Czech Republic,
 CZ-182 21 Praha 8, Czech Republic
\and Center for Computational Materials Science, Technical University,
 A-1060 Vienna, Austria
\and Institute of Physics of Materials, Academy of Sciences of the
 Czech Republic, CZ-616 62 Brno, Czech Republic
\and Max-Planck Institut f\"ur Mikrostrukturphysik, D-06120 Halle, Germany
\and Institut f\"ur Festk\"orperforschung, Forschungszentrum J\"ulich,
 D-52425 J\"ulich, Germany }

\maketitle

\begin{abstract}

{\it Ab initio} formulations of the interlayer exchange
coupling (IEC) between two, in general non-collinearly aligned  magnetic 
slabs embedded in a non-magnetic spacer are reviewed whereby 
both the spacer and the magnetic slabs as well as their interfaces 
may be either ideal or random. 
These formulations are based on the spin-polarized surface Green function 
technique within the tight-binding linear muffin-tin orbital method, 
the Lloyd formulation of the IEC, and the coherent potential approximation
using the vertex-cancellation theorem.
We also present an effective method for the study of the temperature 
dependence of the IEC.
The periods, amplitudes, and phases are studied in terms of discrete 
Fourier transformations, the asymptotic behavior of the IEC is briefly
discussed within the stationary-phase method. 
Numerical results illustrating the theory are presented.

\end{abstract}

\section{Introduction}

Oscillatory interlayer exchange coupling (IEC) has been found 
in a number of ferromagnetic/non-magnetic multilayer systems 
and is in some cases accompanied by an oscillatory magnetoresistance.
The physical origin of such oscillations is attributed to quantum
interferences due to spin-dependent confinement of the electrons in 
the spacer.
The periods of the oscillations with respect to the spacer thickness 
can be correlated to the spacer Fermi surface, a relation frequently
used in experimental studies.
A number of models was proposed to explain this phenomenon and we refer
the reader to excellent recent reviews on the subject\cite{PBr,JMa,IBM}.

The situation is much less satisfactory if the amplitudes and/or phases
are concerned.
They both depend sensitively on the details of the Fermi surface,
and, from the experimental point of view, on the quality
of the multilayers.
Typically, samples include various amounts of disorder at 
interfaces as well as in the bulk (e.g., surface roughness, 
intermixing, impurities, grain boundaries, etc.) which can influence 
the amplitudes and the phases significantly.
 From the theoretical standpoint of view it is important to keep in 
mind that the IEC is an oscillatory phenomenon for which, strictly 
speaking, amplitudes and/or phases are defined only in the
asymptotic limit. 
Experimental data, however, are usually only available for the first 
few oscillations which are sufficient to extract periods, but not 
amplitudes and phases, in particular for the so-called long-period 
oscillations.
The presence of impurities not only complicates the theoretical studies 
but also can provide a valuable insight into the effects controlling
the IEC.
In particular, substitutional alloying can provide a valuable informations 
concerning the topology of alloy Fermi surfaces.
Alloying has also another, more subtle effect, namely it influences
both amplitudes and phases and it can even introduce an extra damping
of the oscillation amplitude (an exponential damping in addition to the 
usual $1/N^{2}$ decay, where $N$ is the spacer thickness) if 
${\bf k}_{\|}$-resolved electron states in the neighborhood of so-called 
callipers (extremal vectors of the Fermi surface) are influenced by 
disorder. 
Finally, we mention that a special case of alloying is intermixing
of magnetic and spacer atoms at interfaces which can significantly
influence coupling amplitudes and which occurs frequently during 
sample preparation in actual experiments.

It is thus obvious that the study of the effect of alloying on the periods, 
amplitudes, and phases of the IEC is an important issue which, however, 
is not properly reflected in the available literature.
Conventional bandstructure methods are of limited use for such
studies although in particular cases, when combined with the 
virtual-crystal-type approximations (VCA), they may be justified,
e.g., for VCr or CrMn alloy spacers studied recently \cite{SHPK}.
However, the complete neglect of alloy disorder makes a reliable
determination of the coupling amplitudes or phases and, to some
extent, even of the coupling periods, uncertain even in such favorable
cases.

In addition, reliable conclusions and verifications of experimental 
measurements can only be based on a parameter-free theory.
In order to determine the IEC one typically estimates the energy 
difference between the ferromagnetic (F) and antiferromagnetic (AF) 
alignment of a system consisting of two magnetic slabs separated by
a non-magnetic spacer.
Using total energy differences (evaluated with the local density 
approximation to the density functional theory) represents an extremely 
difficult task as the tiny exchange energies have to be subtracted 
from the background of huge total energies. Even if one employs very 
fast and accurate linear methods and computational tricks, the spacer 
thickness for which the calculated IEC values are reliable, is limited 
to about 20 layers \cite{ScH,MSAJ}.
On the other hand, for thin spacers this is the most accurate
approach.
One can alternatively employ asymptotic theories which are, strictly 
speaking, valid in the opposite regime, namely, for large spacer and
magnetic slab thicknesses.
The idea is to determine reflection (transmission) coefficients
for an isolated interface between magnetic and spacer metals and
the extremal vectors of the spacer Fermi surface.
The former quantities then determine the coupling amplitudes and phases 
while the latter quantities their periods.
In this case the calculations can be performed by using conventional 
bandstructure methods and, in addition, they will provide a deep insight 
into the physical nature of the IEC \cite{St}.
Note, however, that neither of the above techniques can be extended
to treat disorder nor can they be used to interpolate between two 
limits, namely, the case of thin spacers (preasymptotic region) and 
of thick spacers (asymptotic limit).
For this a theory is needed which can bridge both the preasymptotic 
and the asymptotic region within a unified framework: IEC values for 
a large set of spacer thicknesses (say, for 1-100 atomic layers) can be
analyzed in terms of discrete Fourier transformation in order to reliably
determine not only periods, but also coupling amplitudes and phases.
In addition, one can sample various subsets in order to analyze 
both the preasymptotic and the asymptotic regime as well as 
long-period oscillations.

The basic idea is to determine the IEC directly by employing the 
so-called magnetic force theorem \cite{mft1,mft2} for rotations
in spin space rather then shifting atoms as in the conventional 
force theorem \cite{ft}.
We can thus use the same potentials for both the F and AF (or, in general,
rotated) alignments of the magnetic slabs (the frozen-potential 
approximation) and consider only the single-particle (Kohn-Sham) energies.

This allows a direct formulation of interlayer exchange coupling
based on an application of the Lloyd formula \cite{LlS}
in order to evaluate the difference between 
the grand canonical potentials of the F and AF alignment.
The first calculations of that type were performed by Dederichs's
group in J\"ulich \cite{LNZD}.
The method used in the present paper extends the above approach in three
relevant aspects: 
(i) a reformulation within the framework of a surface Green function 
technique by which linear scaling of the numerical effort with respect 
to the number of layers \cite{iecsgf,iectq} is achieved; 
(ii) a proof of the so-called vertex-cancellation theorem \cite{iecvtx} 
in order to study the influence of alloy disorder on the properties 
of the IEC, and
(iii) an efficient method for a fast and accurate evaluation 
of integrals involving the Fermi-Dirac distribution function in order
to study effects of finite temperature \cite{iectf,iectfa}.
In the present paper we will review these particular techniques that
were developed in the past few years and subsequently applied to a number 
of cases including alloy disorder \cite{iecdb,iecas,iecmor,iecams,iecsor}.
In addition, we have studied systematically the effect of non-magnetic 
cap-layers \cite{ieccap1,ieccap2} on the periods, 
the amplitudes, and the phases of the oscillations of the IEC. 

\section{Formalism}

In this Section we derive an expression for the IEC for in general
non-collinearly aligned  magnetic slabs embedded in a non-magnetic 
spacer.

\subsection{Geometry of the system} \label{sec_geom}
The system considered consists of a stack of layers, namely, from the
left to the right: (i) a semi-infinite (nonmagnetic) substrate, 
(ii) a left ferromagnetic slab of thickness $M$ (in monolayers, MLs), 
(iii) a nonmagnetic spacer of thickness $N$, (iv) a right ferromagnetic 
slab of thickness $M^\prime$, and (v) a semi-infinite (nonmagnetic) 
substrate.
The thickness of the ferromagnetic slabs may extend to infinity.
Eventually, one of the semi-infinite substrates may be substituted by 
a finite nonmagnetic cap of thickness $P$ interfacing semi-infinite 
vacuum.
In general, the various parts of the system can consist of different
metals, including disordered substitutional alloys.
We assume that the spin orientation of the right magnetic slab is rotated
by an angle $\theta$ with respect to that of the left magnetic slab.
In particular, the cases $\theta=0$ and $\theta=\pi$ correspond to
the ferromagnetic and antiferromagnetic alignments of magnetic moments 
of two subsystems, respectively.

\subsection{Electronic structure of the system}

The electronic structure of the multilayer is described by means
of the tight-binding linear-muffin tin orbital (TB-LMTO) method 
\cite{AJ}.
In particular we employ the all-electron scalar-relativistic version
as generalized to the case of random alloys, their surfaces and 
interfaces \cite{book,tkd}.
The key quantity of the formalism, the physical Green function $G(z)$,
is expressed via the auxiliary Green function $g^{\alpha}(z)$ in the
screened tight-binding LMTO representation $\alpha$ as
\begin{equation}
G(z) = \lambda^{\alpha}(z) + \mu^{\alpha}(z) \, g^{\alpha}(z) \,
\mu^{\alpha}(z) \, ,
\label{gftr}
\end{equation}
where
\begin{equation} \label{eq_gf}
 g^{\alpha}(z) = \left( P^{\alpha}(z) - S^{\alpha} \right)^{-1} \, .
\end{equation}
Here $S^{\alpha}$ is a matrix of screened structure constants 
$S^{\alpha}_{{\bf R}L,{\bf R'}L'}$, and $P^{\alpha}(z)$ is a site-diagonal 
matrix of potential functions $P^{\alpha,\sigma}_{{\bf R}L}(z)$.
The potential functions are diagonal with respect to the angular 
momentum index $L=(\ell m)$ and the spin index 
$\sigma=\uparrow,\downarrow$ while the structure constants are 
spin-independent.
The potential functions can be expressed via the so-called
potential parameters $C$, $\Delta$, and $\gamma$ in the following manner
\begin{equation}
P^{\alpha}(z) = \frac{z -C}{\Delta + (\gamma - \alpha)(z-C)} \, ,
\label{pfnc}
\end{equation}
where for matters of simplicity all indices are dropped.
Similarly, the quantities $\lambda^{\alpha}$ and $\mu^{\alpha}$
in~(\ref{gftr}) can be expressed as
\begin{equation}
\lambda^{\alpha}(z) = \frac{\gamma - \alpha}{\Delta +
(\gamma - \alpha)(z-C)} \, , \quad
\mu^{\alpha}(z) = \frac{\sqrt{\Delta}}{\Delta + 
(\gamma - \alpha)(z-C)} \, .
\label{lamu}
\end{equation}
As only the screened representation will be used the superscript $\alpha$
is omitted in the following.

A separate problem is the determination of potential functions $P(z)$
for a given layered structure.
Here we only mention that by employing the magnetic force theorem
we can use the same potential functions for the ferromagnetic and 
rotated (or, antiferromagnetic) alignments.
For random systems treated within the so-called coherent potential 
approximation (CPA) the potential function $P(z)$ is substituted by 
its coherent potential counterpart, ${\cal P}(z)$, whereby the formal
structure of the Green function (\ref{eq_gf}) remains the unchanged.
The methods of determination of (coherent) potential functions for 
collinear alignments of magnetic moments in the present context can be 
found elsewhere \cite{tkd,book}.

\subsection{Definition of the IEC}
The exchange coupling energy ${\cal E}_{x}$, evaluated in the framework
of the magnetic force theorem, is defined as the difference of the 
grand canonical potential 
$\Omega_{\lambda}$ between the ferromagnetic ($\lambda=F$) and 
antiferromagnetic ($\lambda=AF$) alignments of two subsystems, 
i.e.  ${\cal E}_{x} = \Omega_{AF} - \Omega_{F}$.
More generally, the quantity of the physical interest is the difference 
of the grand canonical potentials between a rotated ($\theta \neq 0$) 
and the ferromagnetic  ($\theta=0$) alignment of the two magnetic 
slabs, namely, ${\cal E}_{x}({\theta}) \equiv  \delta \Omega(\theta) = 
\Omega(\theta) - \Omega(0)$.

The grand canonical potential $\Omega$ of a system is defined by
\begin{equation}
\Omega(T,\mu) = - \int_{-\infty}^{\infty}
 f(E,T,\mu) \; N(E) \; {\rm d} E \, ,
\label{omegadef}
\end{equation}
where $N(E)$ is the integrated valence density of states, $f(E,T,\mu)$ 
is the Fermi-Dirac distribution function at the temperature $T$ and the
chemical potential $\mu$ of electrons.
It should be noted that at zero temperature the chemical potential 
coincides with the Fermi energy $E_{F}$ of the system.
The integrated valence density of states is then given by
\begin{equation}
N(E) = - \frac{1}{\pi}\:{\rm Im} \int_{-\infty}^{E}
\: {\rm Tr}\: G(E'+i0) \; {\rm d} E' \, ,
\label{gpmidos}
\end{equation}
where Tr means the trace over lattice sites ${\bf R}$,
angular momentum indices $L=(\ell m)$ and spin indices $\sigma$.
Using~(\ref{pfnc},\ref{lamu}), the following identities can
be verified
\begin{equation}
\frac{d}{d z} \lambda(z) = - \lambda^2(z) \: , \quad
\frac{d}{d z} P(z) =  \mu^2(z) \, .
\label{iden1}
\end{equation}
Together with formula (\ref{deriva}), we find
\begin{equation}
\frac{d}{d z} \Bigl[{\rm Tr\: ln}\: \lambda(z)
+ {\rm Tr\: ln} \: g(z) \Bigr] = - {\rm Tr}\: G(z)  \: .
\label{trick}
\end{equation}
The grandcanonical potential (\ref{omegadef}) is then expressed as
\begin{eqnarray}
\Omega(T,\mu) &=& - \frac{1}{\pi}\:{\rm Im} \int_{-\infty}^{\infty}
f(E,T,\mu) \; {\rm Tr\: ln}\: \lambda(E + i0) \; {\rm d} E
\nonumber \\
&-& \frac{1}{\pi}\:{\rm Im}\int_{-\infty}^{\infty}
f(E,T,\mu) \; {\rm Tr\: ln}\: g(E + i0) \; {\rm d} E \, .
\label{eq_gcpg}
\end{eqnarray}
The formula in (\ref{eq_gcpg}) is the expression for the grandcanonical
potential within the TB-LMTO method \cite{agf} and for finite 
temperatures.

The rotated magnetic configuration is characterized by the set of 
rotation angles
$\Theta=\{\theta_{\bf R}\}$ for all the lattice sites.
In the reference (F) state all the angles $\theta_{\bf R}=0$ 
while in the rotated state  $\theta_{\bf R}=\theta$ in the rotated
magnetic layer and $\theta_{\bf R}=0$ for all other lattice sites.
The quantities $\lambda(\Theta,z)$ and $g(\Theta,z)$ for the rotated
system are given by
\begin{equation}
\lambda(\Theta,z)={\sf U}(\Theta)\lambda(0,z){\sf U}^{\dagger}(\Theta)
\, , \;
g(\Theta,z)=[{\sf U}(\Theta)P(0,z){\sf U}^{\dagger}(\Theta)-S]^{-1} \, .
\label{rottheta}
\end{equation}
Here $[{\sf U}(\Theta)]_{\bf RR'} = \delta_{\bf RR'} U(\theta_{\bf R})$
is the rotation matrix for spin 1/2 particles defined in terms of the 
single-site matrices $U(\theta_{\bf R})$ \cite{AHII}
\begin{equation}
{\rm U(\theta)} = \left(
\begin{array}{cc}
\ c & \quad s \\
-s  & \quad c
\end{array} 
\right) \, ,
\label{umat}
\end{equation}
where $c=\cos(\theta/2)$, $s=\sin(\theta/2)$,
${\rm U}(\theta) \, {\rm U}^{\dagger}(\theta)$=
${\rm U}^{\dagger}(\theta) \, {\rm U}(\theta)= 1$, and
${\rm det U}(\theta)={\rm det U}^{\dagger}(\theta)= 1$.
We note that in the rotated magnetic configuration
$P(\Theta,z)={\sf U}(\Theta)P(0,z){\sf U}^{\dagger}(\Theta)$ is
generally a non-diagonal matrix with respect to the spin
indices $\sigma,\sigma'$.

The first term in (\ref{eq_gcpg}) is independent of $\theta$ because
$\lambda(z)$ is site (and layer-) diagonal, it therefore does not 
contribute to the exchange energy ${\cal E}_{x}({\theta})$, i.e.,
it is sufficient to consider the second part only,
\begin{equation}
\Omega(\theta,T,\mu) = 
-\frac{1}{\pi}\:{\rm Im}\int_{-\infty}^{\infty} f(E,T,\mu)
\: {\rm Tr\: ln}\: g(\theta,E + i0) \; {\rm d} E \, .
\label{eq_gcp}
\end{equation}
It should be noted that the above expression is valid only in the 
absence of spin-orbit coupling.  

The magnetic force theorem used here for the evaluation of the IEC was 
used also in related problems, e.g., for the evaluation of the
exchange energies of two impurities embedded in a nonmagnetic host
\cite{mft1} and then extended to the case of Heisenberg exchange
parameters between two sites in a magnetic material\cite{mft2}.
In the latter case the magnetic force theorem is valid only for the
infinitesimal rotations while in the former case it is valid also for 
$\theta=\pi$ \cite{LNWZD}.

\subsection{Configurational averaging}
Keeping in mind applications to random systems, one is interested
in the configurational average of the expression in (\ref{eq_gcp}),
namely,
\begin{equation} \label{eq_agcp}
\left< \Omega \right> = - \frac{1}{\pi} \, {\rm Im}
\int_{-\infty}^{\infty} f(E,T,\mu) \; \langle {\rm Tr \;  ln} \, 
g(E + i0) \, \rangle \; {\rm d} E \, ,
\label{eq_confav}
\end{equation}
where $\langle \dots \rangle$ denotes a configurational average.
Difficulties here arise from the fact that the configurational average 
of the logarithm $\left< \ln \, g(z) \right>$ can differ
significantly from the logarithm of the configuration average $\ln
\left< g(z) \right>$.  
The difference 
$\mbox{\sf X} \equiv \left< \ln \, g \right> - 
\ln \left< g \right>$, the so-called vertex correction, 
is difficult to calculate and usually cannot be neglected.  
Fortunately, this problem can be circumvented by using the vertex 
cancellation theorem \cite{iecvtx}, which states that the contributions 
from the vertex correction for the $F$ and $AF$ configurations cancel 
each other exactly, namely 
$\mbox{Tr \sf X}_{AF} - \mbox{Tr \sf X}_{F} = 0$, such that to first
order with respect to the angle between the magnetizations in the 
two ferromagnetic layers vertex corrections can be omitted.
In other words, the evaluation of (\ref{eq_confav}) simplifies to
\begin{eqnarray} 
\left< \Omega \right> &=& - \frac{1}{\pi} \, {\rm Im}
\int_{-\infty}^{\infty} f(E,T,\mu) \; {\rm Tr \;  ln} \,
\langle g(E + i0) \, \rangle \; {\rm d} E \, ,
\nonumber  \\
&=& - \frac{1}{\pi} \, {\rm Im}
\int_{C} f(z,T,\mu) \; {\rm Tr \;  ln} \,
\langle g(z) \, \rangle \; {\rm d} z \, . 
\label{eq_vtxcan}
\end{eqnarray}
We have also substituted the energy integral by integration over 
a contour in the complex energy plane $z$.
The possibility to neglect vertex corrections can conveniently be used 
in calculations of the interlayer exchange coupling as explicit 
numerical calculations have shown that it remains valid to a good accuracy 
even for an angle as large as $\pi$ \cite{iecvtx}.
In this respect it is very similar to the force theorem \cite{ft}.  
It is important to note that such an extension is only applicable to 
the evaluation of exchange energies of magnetic systems 
interacting via a non-magnetic host.
An evaluation of exchange energies in ferromagnetic systems such as
parameters of a classical Heisenberg model, was claimed to be limited to 
infinitesimal rotations only \cite{mft2}. 
The use of the vertex-cancellation theorem allows to reduce the computational 
time in first-principles calculations by almost two orders of magnitude, 
so that the computational effort for disordered systems is comparable 
to that for a pure system \cite{iecvtx}.
We refer the reader to Appendix A for more details concerning the 
derivation and applicability of the vertex-cancellation theorem.
The last remark concerns the fact that the expression for the change
in the grandcanonical potential within the magnetic force theorem
also includes the classical magnetostatic dipole-dipole interaction 
energy (DDIE). 
The DDIE decays with a spacer thickness much faster than the IEC and its 
contribution can be thus neglected for thicker spacer anyhow.
In addition, first-principles fully-relativistic calculations of the
IEC \cite{SUWS} have demonstrated that this term has a negligible
influence even for a rather thin spacer amounting just to a few layers.
Consequently, the DDIE term will be neglected in the following.  

\subsection{Lloyd formula}
We need to evaluate the difference of configurationally averaged
grandcanonical potentials in the rotated and FM configurations.
This can be done conveniently with the help of the well-known Lloyd
formula \cite{LlS} applied to layered systems.
We formally split the system into two non-interacting fragments,
namely a left fragment ${\cal L}$, which consists of the left substrate
and the left magnetic slab, and a right fragment ${\cal R}$, which 
comprises the rest of the system, i.e., the spacer, the right 
magnetic slab, and the right substrate (or, eventually, the cap layer
interfacing the vacuum).
Fragments are described by the unperturbed Green function 
$\left< g_{0} (z) \right>$.
In the next step we couple two fragments together with help of a
localized potential $V$ which is simply the interlayer screened 
structure constant.
This procedure has a number of advantages as compared to a
conventional way of embedding two finite magnetic layers into the
infinite (bulk) host spacer \cite{LNZD}: (i) the perturbation $V$
is independent of the thicknesses of magnetic layers; (ii) complicated
sample geometries can be treated, including semi-infinite magnetic layers; 
and (iii) a powerful and efficient method exists for the evaluation of 
the Green function of fragments, namely the surface Green function
technique in the principal-layer formulation \cite{book,tkd}.

Keeping in mind the vertex cancellation theorem, one gets for a
difference in the configurationally averaged grandcanonical 
potential~(\ref{eq_vtxcan}), the expression
\begin{equation} \label{eq_Lloyd}
\left< \delta \Omega \right> = - \frac{1}{\pi} \, {\rm Im}
\int_{C} f(z,T,\mu) \; {\rm Tr \,  ln} \, 
\left( 1 - V \left< g_{0}(z) \right> \right) \, {\rm d} z \, ,
\end{equation}
where $\left< g_{0} (z) \right>$ is the configurationally 
averaged Green function of the decoupled non-interacting fragments 
${\cal L}$ and ${\cal R}$ defined above.
For the sake of simplicity, we will denote from here on the 
configurationally averaged quantities by an overbar, e.g., 
$\left< g_{0} (z) \right> \equiv {\bar g}_{0}(z) $.
The concept of principal layers (PL) \cite{WKu} as used within
the TB-LMTO method leads to a block tridiagonal form of the 
structure constants and of the inverse Green function.
If we apply this tridiagonality to~(\ref{eq_Lloyd}), we get for
$V$ and $\left< g_{0} (z) \right>$ the following expressions 
by using a supermatrix notation with respect to nearest-neighbor PLs
resolved in the wave-vector ${\bf k}_{\|}$,
\begin{equation} \label{eq_vg}
V({\bf k}_{\|})= \left(
\begin{array}{cc}
0 &  
S_{10}({\bf k}_{\|}) \\
S_{01}({\bf k}_{\|}) &  
0
\end{array}
\right) \, , \;\;
\bar{g}_{0} ({\bf k}_{\|},z) = \left(
\begin{array}{cc}
\bar{{\cal G}}_{\cal L} ({\bf k}_{\|},z)  & 0 \\
0 & \bar{{\cal G}}_{\cal R} ({\bf k}_{\|},z) 
\end{array}
\right) \, ,
\end{equation}
where $S_{10}({\bf k}_{\|})=\left[ S_{01}({\bf k}_{\|}) \right]^{\dagger}$.
Combining~(\ref{eq_Lloyd}) and (\ref{eq_vg}) one gets 
\begin{eqnarray} \label{eq_iec1}
\delta \, {\rm Tr \; ln} \, {\bar {\cal G}} (z)  &=&
- \frac{1}{N_{\|}} \sum_{{\bf k}_{\|}} \, {\rm tr \; ln} \,
\left[ 1 - {\bar \Gamma}_{\cal L}({\bf k}_{\|},z) \,
\bar{{\cal G}}_{\cal R}({\bf k}_{\|},z) \right] \, ,
\\ \nonumber
{\bar \Gamma}_{\cal L}({\bf k}_{\|},z) &=&
S_{10}({\bf k}_{\|}) \, 
\bar{{\cal G}}_{\cal L}({\bf k}_{\|},z) \, 
S_{01}({\bf k}_{\|}) \, .
\end{eqnarray}
Here the quantity ${\bar \Gamma}_{\cal L}({\bf k}_{\|},z)$ has the 
meaning of an effective embedding potential, and the quantities 
${\bar {\cal G}_{\cal L}}$ and ${\bar {\cal G}_{\cal R}}$ are 
the configurationally averaged surface Green functions (SGF) \cite{WKu} 
of the magnetic subsystems ${\cal L}$ and ${\cal R}$, respectively.
By definition, the surface Green function ${\bar {\cal G}_{\cal S}}$
(${\cal S=L,R}$) is the top PL projection of the Green function 
of the corresponding semi-infinite system ${\cal S}$.
Its determination in the case of random systems was extensively 
discussed in the literature, see \cite{sgf1,sgf2,sgf3,book}.
The summation in~(\ref{eq_iec1}) extends over the surface
Brillouin zone (SBZ) corresponding to the underlying two-dimensional 
translational symmetry \cite{pwsymm}, and $N_{\|}$ is the number of 
sites in a layer. 

\subsection{The IEC for a general angle $\theta$}
Let us now turn to the evaluation of the
energy difference between  arbitrary alignments.
Consider the following quantity,
\begin{eqnarray}
{\rm tr \, ln \, Z} = {\rm tr \, ln \,(1 - A_{0} \ B) -
                     tr \, ln \, (1 -A_{0} \ B_{0})} \, ,
\label{difz}
\end{eqnarray}
where the matrices A$_0$ and B$_0$ are related to the ferromagnetic 
alignment and thus are diagonal in spin space
\begin{equation}
{\rm A}_{0} = \left(
\begin{array}{cc}
{\rm A}_0^{\uparrow} & 0 \\
0  & {\rm A}_0^{\downarrow}
\end{array}
\right) \, , \quad
{\rm B}_{0} = \left(
\begin{array}{cc}
{\rm B}_0^{\uparrow} & 0 \\
0  & {\rm B}_0^{\downarrow}
\end{array}
\right) \, . 
\label{a0b0}
\end{equation}
The particular form of the subblocks A$^{\sigma}_0$ and 
B$^{\sigma}_0$ ($\sigma= \uparrow, \downarrow$) is given by
\begin{equation} \label{eq_AB}
{\rm A_{0}^{\sigma}}=S_{10}({\bf k}_{\|}) \,
{\bar {\cal G}_{\cal L}}^{\sigma}({\bf k}_{\|},z) \, 
S_{01}({\bf k}_{\|}) \, , \;\;\;\;\; 
{\rm B_{0}^{\sigma}}=
{\bar {\cal G}_{\cal R}}^{\sigma}({\bf k}_{\|},z) \, .
\end{equation}

The matrix B refers to an alignment in which the orientations of 
the magnetization in two magnetic slabs are rotated uniformly by
a relative angle $\theta$,
\begin{equation}
{\rm B} = {\rm U}(\theta) \, {\rm B}_0 \, {\rm U}^{\dagger}(\theta) 
\, ,
\end{equation}
where ${\rm U(\theta)}$ is the rotation matrix (\ref{umat}).
The quantity $1 - {\rm A_{0} \, B}$ in (\ref{difz}) can therefore be 
written as
\begin{equation}
{\rm 1 - A_{0} \ B} = \biggl({\rm U}(\theta) - 
{\rm A}_0 {\rm U}(\theta) {\rm B}_0\biggr) 
{\rm U}^{\dagger}(\theta) \, ,
\label{aubu}
\end{equation}
where, as follows from~(\ref{a0b0}) and (\ref{umat}),
\begin{equation}
{\rm U(\theta) - A_{0} \ U(\theta) \ B_{0}} = \left(
\begin{array}{cc}
c \rm  \ (1 - A^{\uparrow}_{0} \ B^{\uparrow}_{0})
& \quad s \rm \ (1 - A^{\uparrow}_{0} \ B^{\downarrow}_{0})\\ 
-s \rm \ (1 - A^{\downarrow}_{0} \ B^{\uparrow}_{0}) 
& \quad c \rm \ (1 - A^{\downarrow}_{0} \ B^{\downarrow}_{0})
\end{array}
\right) \, .
\end{equation}
Using now the identity ${\rm tr \, ln \ X}$ = ${\rm ln \, det \ X}$, 
which is valid for any non-singular matrix X, and the identity
\begin{equation}
\rm det
\left(
\begin{array}{cc}
{\rm A} & {\rm B} \\ \, {\rm C}  & \, {\rm D}
\end{array}
\right) = 
det A \, . \, det D \, . \, det (1 - A^{-1} \, B \, D^{-1} \, C) \, ,
\label{eq_lnz}
\end{equation}
which in turn is valid, if the matrices A and D are non-singular, 
it is straightforward to prove that
\begin{equation}
\rm tr \, ln \ Z = tr_{\it L} \, 
ln \, \left(1 - \frac{1 - \cos (\theta)}{2} \, M \right) \, ,
\end{equation}
where
\begin{equation} \label{eq_M}
{\rm M = 1 - (1 - A_{0}^{\uparrow} \, B_{0}^{\uparrow})^{-1} \;
(1 - A_{0}^{\uparrow} \, B_{0}^{\downarrow}) \,
(1 - A_{0}^{\downarrow} \, B_{0}^{\downarrow})^{-1} \;
(1 - A_{0}^{\downarrow} \, B_{0}^{\uparrow})} \, .
\end{equation}
It should be noted that in (\ref{difz}) tr denotes the trace over 
angular momenta and spin, while in (\ref{eq_lnz}) tr$_L$ denotes the 
trace over orbital momenta only.
The final expression for ${\cal E}_{x}(\theta)$ is thus given by
\begin{eqnarray} \label{eq_iec}
{\cal E}_{x}({\theta}) = \frac{1}{\pi N_{\|}} \sum_{{\bf k}_{\|}} \,
{\rm Im} \, \int_{C} & & f(z,T,\mu) \times \nonumber \\
& & {\rm tr}_L \,{\rm ln} \,
\left(1 - \frac{1 - \cos (\theta)}{2} \, {\rm M} ({\bf k}_{\|},z) \right) 
\, {\rm d} z \, ,
\end{eqnarray}
in which the energy integral is expressed in terms of a contour 
integral which will be discussed in detail later.

It is interesting to note that the expression~(\ref{eq_M}) for 
${\rm M}({\bf k}_{\|},z)$ can be rearranged in the
following form \cite{iecas}
\begin{eqnarray} \label{eq_M1}
{\rm M} = - & & \left( 1 - S_{10} \, {\bar {\cal G}}_{\cal L}^\uparrow \,
S_{01} \, {\bar {\cal G}}_{\cal R}^\uparrow \right)^{-1}
S_{10} \left( {\bar {\cal G}}_{\cal L}^\uparrow -
{\bar {\cal G}}_{\cal L}^\downarrow \right) \times \nonumber \\
& & \left( 1 - S_{01} \,
{\bar {\cal G}}_{\cal R}^\downarrow \,
S_{10} \, {\bar {\cal G}}_{\cal L}^\downarrow \right)^{-1}
S_{01} \left( {\bar {\cal G}}_{\cal R}^\uparrow -
{\bar {\cal G}}_{\cal R}^\downarrow \right) \, .
\end{eqnarray}
It explicitly factorizes the 'spin-asymmetry' of the problem and 
it is directly related to RKKY-like theories \cite{PBr}.
This result \cite{iecas} is formally equivalent
to the results of the spin current approach \cite{sloncz}
as formulated within a Green function formalism based on an 
empirical single orbital tight-binding model \cite{iecsc}. 
A matrix version developed in the framework of a semiempirical 
tight-binding model has appeared recently \cite{JMa}.

For completeness we also give the result for the common case of the
antiferromagnetic alignment ($\theta=\pi$):
\begin{equation} \label{eq_iecpi}
{\cal E}_{x} \equiv {\cal E}_{x}({\pi}) = 
\frac{1}{\pi N_{\|}} \sum_{{\bf k}_{\|}} \, {\rm Im} \, \int_{C} 
f(z,T,\mu) \; {\rm tr}_L \,{\rm ln} \, {\cal M} ({\bf k}_{\|},z) \, 
{\rm d} z \, ,
\end{equation}
where ${\cal M}$ is a product of four terms,
\begin{equation} \label{eq_Mpi}
{\cal M} = {\rm (1 - A_{0}^{\uparrow} \, B_{0}^{\uparrow})^{-1} \;
(1 - A_{0}^{\uparrow} \, B_{0}^{\downarrow}) \,
(1 - A_{0}^{\downarrow} \, B_{0}^{\downarrow})^{-1} \;
(1 - A_{0}^{\downarrow} \, B_{0}^{\uparrow})} \, .
\end{equation}

\subsection{The torque and infinitesimal rotations}
The differential change in the grand canonical potential 
$\delta \Omega(\theta)$ with respect to a differential relative
angle $\theta$, 
$ -\partial \, \delta \Omega(\theta)/ \partial \, \theta$,
is usually called the torque.
The torque can easily be obtained by
differentiating~(\ref{eq_iec}) with respect to the angle $\theta$.
By definition one gets therefore
\begin{equation}
T({\theta}) = -\frac{\partial {\cal E}_{x}({\theta})}
{\partial {\theta}} \, \;\;\;\;\;\; {\rm or} \;\;\;\;\;\;
{\cal E}_{x}({\theta}) = - \int_{0}^{\theta} \, T(\theta')
\, d{\theta'} \, ,
\end{equation}
whereby $T({\theta})$ follows immediately from (\ref{eq_iec})
\begin{eqnarray}
T({\theta}) = \frac{\sin({\theta})}{2 \pi N_{\|}} & & \sum_{{\bf k}_{\|}}
\, {\rm Im} \, \int_{C} f(z,T,\mu) \times \nonumber \\
& & {\rm tr}_L \, \left[ {\rm M}({\bf k}_{\|},z) \, 
\left( 1 - {1 \over 2} \, [ 1 - \cos ({\theta})] \,
{\rm M}({\bf k}_{\|},z) \right)^{-1} \right] \, {\rm d} z \, .
\end{eqnarray}
By formally expanding the logarithm in (\ref{eq_iec}) 
in powers of ${\rm 1 - cos}(\theta)$, one can cast the expression for 
${\cal E}_{x}(\theta)$ into the form
\begin{equation} \label{eq_expan}
{\cal E}_{x}({\theta}) = B_{1} \, [ 1 - \cos(\theta)] + 
\frac{1}{2} \, B_{2} \, [ 1 - \cos(\theta)]^{2} + \, \dots \, ,
\end{equation}
where $B_{1}$ and $B_{2}$ are the so-called bilinear and the (intrinsic)
biquadratic exchange coupling coefficients, respectively,
\begin{eqnarray}
B_{1} &=& \frac{1}{2 \pi  N_{\|}} \sum_{{\bf k}_{\|}} \, {\rm Im}
\int_{C} f(z,T,\mu) \, {\rm tr}_L \;  {\rm M}({\bf k}_{\|},z) 
\, {\rm d} z \, , \\ \nonumber
B_{2} &=& - \frac{1}{4 \pi \, N_{\|}} \sum_{{\bf k}_{\|}} \, 
{\rm Im} \int_{C} f(z,T,\mu) \, {\rm tr}_L \;  
[ {\rm M}({\bf k}_{\|},z) ]^{2} \, {\rm d} z \, .
\end{eqnarray}
It may be, however, more convenient to fit the exact
expression~(\ref{eq_iec}) into the form (\ref{eq_expan}) by employing  
calculated values for $\theta=\pi/2$ and $\theta=\pi$ \cite{gmr}.
We obtain
\begin{equation} 
B_{1}= \frac{{\cal E}_{x}({\pi}) + 2 {\cal E}_{x}({\pi/2)}}{2} \, ,
\;\;\;\;\;\;
B_{2}= \frac{{\cal E}_{x}({\pi}) - 2 {\cal E}_{x}({\pi/2)}}{2} \, .
\end{equation}

Of particular interest is the expansion of ${\cal E}_{x}({\theta})$ 
for a small $\theta$, i.e., when ${\rm 1-cos}(\theta)$ is a small 
parameter (the method of infinitesimal rotations (MIR)).
This approach becomes particularly relevant in the case when the spacer 
is a magnetic metal or for complicated geometries, e.g., for
so-called periodic multilayers.

\subsection{The IEC as interface-interface interaction}
\label{intint}
We will now discuss briefly an alternative approach of a direct 
evaluation of the IEC as a difference in the interface-interface
interaction energies rather then its indirect determination in terms of 
the energy of a single interface~(\ref{eq_agcp}-\ref{eq_vg}).
We decouple the system into three fragments, a left, central, and
right fragment, $\cal L$, $\cal C$, and $\cal R$, respectively.
The left and the right fragment are formed by corresponding substrates 
with magnetic slabs whereby the central slab comprises the spacer.
Both approaches are physically equivalent because it is irrelevant how
the system is divided into an unperturbed part and a perturbation.
Note, however, that the interface-interface formulation is more general
as it could be used for a determination of interaction energies of two
generally different interfaces.

The derivation proceeds in two steps and employs partitioning 
technique with respect to the trace of the logarithm of the Green 
function.
First, the subsystems ${\cal L}$ and ${\cal R}$ are downfolded which 
leads to an effective problem of two localized perturbations in the 
subsystem ${\cal C}$. 
The second step, a two-potential formula applied to the fragment
${\cal C}$ separates directly the interface-interface contribution.
The result has formally the same structure as the previous
one~(\ref{eq_iec1},\ref{eq_M}), but the subblocks A$^{\sigma}_0$ and 
B$^{\sigma}_0$ ($\sigma=\uparrow, \downarrow$) are now of the following 
form \begin{equation} \label{eq_iec2}
{\rm A_{0}^{\sigma}}={\bar g}_{N1}({\bf k}_{\|},z) \,
\bar{\tau}_{1}^{\sigma}({\bf k}_{\|},z) \, 
{\bar g}_{1N}({\bf k}_{\|},z) \, ,
\;\;\;\;\; {\rm B_{0}^{\sigma}}=
\bar{\tau}_{N}^{\sigma}({\bf k}_{\|},z) \, .
\end{equation}
The $\tau$-matrices ${\bar \tau}_{i}$ ($i=1,N$) corresponding to 
"multiple scattering" at individual interfaces ${\cal {L/C}}$, ($i=1$) 
and ${\cal {C/R}}$, ($i=N$) are expressed as
\begin{equation}
{\bar \tau}_{i}^{\sigma}({\bf k}_{\|},z) =
{\bar \Gamma}_{i}^{\sigma}({\bf k}_{\|},z) \,
\left[ 1 - {\bar g}_{ii}({\bf k}_{\|},z) \,
{\bar \Gamma}_{i}^{\sigma}({\bf k}_{\|},z) \right]^{-1} \, ,
\end{equation}
where the effective embedding potentials 
${\bar \Gamma}_{i}^{\sigma}({\bf k}_{\|},z)$ of the left and right
interfaces ($i=1,N$), respectively, are defined as
\begin{eqnarray} \label{eq_iec2a}
{\bar \Gamma}_{1}^{\sigma}({\bf k}_{\|},z) = S_{10}({\bf k}_{\|})
\, {\bar {\cal G}}_{\cal L}^{\sigma}({\bf k}_{\|},z)
\, S_{01}({\bf k}_{\|}) \, , \\
\nonumber
{\bar \Gamma}_{N}^{\sigma}({\bf k}_{\|},z) = S_{01}({\bf k}_{\|})
\, {\bar {\cal G}}_{\cal R}^{\sigma}({\bf k}_{\|},z)
\, S_{10}({\bf k}_{\|}) \, .
\end{eqnarray}
Here, ${\bar {\cal G}}_{\cal S}^{\sigma}$ (${\cal S=L,R}$) are 
the configurationally averaged SGFs of the left and the right 
semi-infinite regions, respectively. 
Details of the derivation can be found in Appendices B and C.
The coupling between the two magnetic subsystems is due to
the layer off-diagonal projections ${\bar g}_{1N}({\bf k}_{\|},z)$ 
and ${\bar g}_{N1}({\bf k}_{\|},z)$ of the Green function (GF) of 
the finite spacer consisting of $N$ layers.  
The oscillatory behavior of interlayer coupling is then
governed by the oscillatory behavior of these quasi
one-dimensional spacer Green functions, a formulation which is
very much in the spirit of a simplified RKKY approach \cite{PBr}.
An efficient method of evaluation of the corner-blocks of the Green 
function, ${\bar g}_{ij}({\bf k}_{\|},z)$, $(i,j=1,N)$, is
described in Appendix D \cite{Godf,sgf3}.

\subsection{Relation to the KKR method}
We shall discuss now the relation of the present 
approach~(\ref{eq_iecpi},\ref{eq_Mpi},\ref{eq_iec2}) to the method employed 
in \cite{LNZD} and based on the Korringa-Kohn-Rostoker (KKR) 
Green function technique.
Let us note first the deep internal connection between the KKR and
the TB-LMTO-GF approach (see~\cite{book,WTS} for more details). 
The model in (\ref{eq_gcp}) consists of an infinite ideal non-magnetic 
spacer as a reference system and of two magnetic slabs representing 
localized perturbations.
For simplicity we start from the case of two magnetic monolayers in 
an infinite spacer.
The result 
\begin{equation} \label{eq_iecdede}
{\rm A_{0}^{\sigma}}={\cal G}_{N1}^{b}(z)({\bf k}_{\|},z) \,
t_{1}^{\sigma}({\bf k}_{\|},z) \, {\cal G}_{1N}^{b}({\bf k}_{\|},z) 
\, , \;\;\;\;\; {\rm B_{0}^{\sigma}}=
t_{N}^{\sigma}({\bf k}_{\|},z) 
\end{equation}
is formally the same with the exception that the 
$\tau$-matrices entering~(\ref{eq_iec2}) are now substituted by
the single-site t-matrices $t_{i}$ which describe the scattering of 
electrons from two magnetic monolayers at $i=1,N$ embedded in an infinite 
non-random bulk spacer and separated by $N-2$ spacer layers:
\begin{equation} \label{eq_tm}
t_{\lambda;\, i}^{\sigma}({\bf k}_{\|},z) =
\Delta P_{\lambda;\, i}^{\sigma}(z) \,
\left[ 1 + {\cal G}^{b}({\bf k}_{\|},z) \,
\Delta P_{\lambda;\, i}^{\sigma}(z) \right]^{-1} \, .
\end{equation}
The strength of the scattering potential,
$\Delta P_{\lambda;\, i}^{\sigma}(z)$, is given by the difference of
the potential functions for the magnetic monolayer
$P^{\sigma}_{\lambda;\, i}(z)$ and for the non-magnetic spacer
$P(z)$, while ${\cal G}^{b}({\bf k}_{\|},z)$ is the layer diagonal
block of the GF of the bulk spacer.
The layer off-diagonal blocks of the bulk spacer GF,
${\cal G}_{1N}^{b}(z)$ and ${\cal G}_{N1}^{b}(z)$, are given by
\begin{equation} \label{eq_bgf}
{\cal G}_{1N}^{b}({\bf k}_{\|},z) =
\left[ {\cal G}^{s}({\bf k}_{\|},z) \, S_{01}({\bf k}_{\|})
\right]^{N-1} \, {\cal G}^{b}({\bf k}_{\|},z) \, ,
\end{equation}
and similarly for ${\cal G}_{N1}^{b}(z)$.
Here, ${\cal G}^{s}({\bf k}_{\|},z)$ is the corresponding SGF
of an ideal semi-infinite non-magnetic bulk spacer \cite{WKu}.
It should be noted that also the layer-resolved bulk Green
function ${\cal G}^{b}({\bf k}_{\|},z)$ can be expressed in
terms of the SGFs (see, e.g.,~\cite{sgf1}).
Since (\ref{eq_bgf}) is exact, there is no need to perform an additional
$k_{\perp}$-integration \cite{LNZD}.
It is easy to show that the result is formally identical to
the case of two impurities in a simple tight-binding linear chain 
model with nearest neighbor hopping.

A generalization to the case of magnetic slabs containing a 
finite number M of magnetic layers is formally straightforward 
\cite{LNZD}.
The t-matrices $t^{\sigma}_{\lambda; i}(z)$ are then
supermatrices with respect to angular momentum and layer
indices and the numerical effort to evaluate~(\ref{eq_tm}) increases
with the third power of M as contrasted with the results of the present
approach~(\ref{eq_iec1},\ref{eq_iec2}) which depend only linearly 
on M.

\subsection{Influence of external periodicity}
\label{sc_ext}

Until now it was assumed implicitly that we have a simple "parent"
lattice \cite{pwsymm}.
The periods of the coupling oscillations are closely related to the 
Fermi surface geometry \cite{PBr,JMa} of the bulk spacer.
A different translational symmetry (complex lattices) or stacking
sequence within layers will thus tend for sufficiently thick spacers 
to a different kind of bulk periodicity and hence to new periods.
For example, an alternating stacking of fcc(001)-layers Cu and ordered 
c($2 \times 2$)-CuAu layers tends to an ordered fcc-Cu$_{3}$Au alloy 
with a Fermi surface topology different from that of fcc-Cu spacer.
For a discussion of "superlattice" formation in magnetic multilayers
see also~\cite{sform}.
We will discuss in the following in some detail two possibilities, 
namely superstructures in the spacer and in the magnetic slabs.

We start with the former case by assuming the same geometry as discussed
in Sec.~\ref{sec_geom} but now the spacer slab consists
of two non-magnetic metals $A$ and $B$ with respective thickness $n_{A}$
and  $n_{B}$ periodically alternating.
Typically, the spacer layer starts with the layer $A (B)$ and ends with 
the layer $B (A)$, but the termination of the spacer slab with the same
layers is also possible (and interesting \cite{iecsor}). 
The particular case of $n_{A}=n_{B}=1$ 
corresponding to an (001)-stacking of an ordered fcc-CuAu alloy was 
already treated on a first-principles level \cite{iecsor}.
The more general case, ($n_{A}, n_{B} > 1$), which corresponds to
artificially grown superstructures, was treated only within a simple 
one-band model \cite{iecsor1}.
In both cases, new periodicities (in comparison with the spacers
consisting from pure $A$ or $B$ metals) arise with an increasing
number of repetitions.
Alternatively, one can consider a superstructure within a given spacer 
layer, or combination of both, e.g., the above mentioned example of the 
ordered fcc-Cu$_{3}$Au alloy spacer. 
The similar situation can be encountered also in the magnetic slabs.
In particular the case of a c($2 \times 2$)-CoFe periodicity
within the magnetic layers separated by a fcc-Cu(001) spacer \cite{iecmor} 
leads to the rather surprising appearance of new periods.
These new periods can be now correlated to critical points
of the spacer Fermi surface folded down to the Brillouin zone
corresponding to a c(2$\times$2)-superlattice \cite{iecmor}.
A correlated gradual appearance of new periods and the order 
in statistically disordered layers is a clear indication of
their relation to a different bulk periodicity \cite{iecmor,iecsor}.

A special case of alternating layers of $A$ and $B$ metals is when one 
of metals is magnetic and the other is nonmagnetic, all of which 
sandwiched between two substrates.
This is the case of a periodic multilayer.  

The generalization of the present formalism to above discussed cases 
is rather straightforward.
In the case of a superlattice within a layer it is just sufficient to
substitute matrices appearing in~(\ref{eq_iec},\ref{eq_iecpi}) by the
corresponding supermatrices, e.g., by ($2 \times 2$)-supermatrices in 
the case of a c($2 \times 2$)-superlattice. 
The key quantity, the surface Green functions 
${\bar {\cal G}}^{\sigma}_{\cal L,R}$~(\ref{eq_AB}), can be
easily evaluated also in this case (see for details~\cite{book}).
The generalization of the formalism to the case of alternating layers
from $A$ and $B$ metals is as well simple because the surface Green
function is constructed in an epitaxial manner, i.e., layer by layer, 
and it is therefore immaterial if the stacking of layers consists 
of the same or a different material. 
In the limit of a periodic multilayer we should just keep in mind that 
a proper repeating unit consists now from four layers, namely $S-M-S-M$, 
where the symbols $S$ and $M$ refer to the spacer and magnetic 
layers, respectively.
This is necessary to calculate the F and AF configurations needed for
the evaluation of the IEC. 
We note that the present formalism allows to evaluate efficiently and
reliably the IEC for thick spacers (one hundred layers and more) which 
is important for realistic studies of so-called superlattice spacers 
and of periodic multilayers.
  
\subsection{Temperature-dependence of the IEC}
We conclude this Section by reviewing a recently developed technique 
for an efficient evaluation of the temperature dependence of the IEC
\cite{iectf}.
The main cause for the temperature dependence of the IEC is connected
with thermal excitations of electron-hole pairs across the Fermi level
as described by the Fermi-Dirac function.
It turns out that other mechanisms (as for example electron-phonon 
and electron-magnon interactions) are less important.
We rewrite~(\ref{eq_iecpi}) into the following form
\begin{eqnarray} \label{eq_iecT}
{\cal E}_x(T) &=& {\rm Im} \, I(T) \, , \quad
I(T) = \int_{C} f(z,T,\mu) \, \Psi(z) \, {\rm d} z \, ,
\end{eqnarray}
where 
\begin{eqnarray} \label{eq_psi}
\Psi(z) &=& \frac{1}{\pi N_{\|}} \, \sum_{{\bf k}_{\|}} \,
{\rm tr}_{L} \, {\rm ln} \, {\cal M}({\bf k}_{\|},z) \, ,
\end{eqnarray}
with the energy integration performed over a contour $C$ along the real
axis and closed by a large semicircle in the upper half of the complex
energy plane.

The integral in (\ref{eq_iecT}) can be recast into a more suitable form
using the analytic properties of $\Psi(z)$, namely, (i) $\Psi(z)$
is holomorphic in the upper half of the complex halfplane, and (ii)
$z \Psi(z) \rightarrow 0$ for $z \rightarrow \infty, \, {\rm Im} z > 0$.
Let us define a new function $\Phi(y) = -i \, \Psi(E_F+iy)$ of a real
variable $y$, $y \geq 0$.
Then at $T=0$ K,
\begin{eqnarray} \label{eq_I0}
I(0) = \int_{0}^{+\infty} \Phi(y) \, {\rm d} y \, ,
\end{eqnarray}
while at $T>0$ K,
\begin{eqnarray} \label{eq_IT}
I(T) = 2 \pi k_B T \sum_{k=1}^{\infty} \Phi(y_k) \, ,
\end{eqnarray}
where $k_B$ is the Boltzmann constant and the $y_k$ are Matsubara 
energies, $y_k = \pi k_B T (2k - 1)$.
In the limit $T \rightarrow 0$, $I(T) \rightarrow I(0)$ continuously.

We have verified that the function $\Phi(y)$ can be represented
accurately as a sum of a few complex exponentials of the form
\begin{equation} \label{eq_pce}
\Phi(y)=\sum_{j=1}^M \, A_j \, {\rm exp} (p_j y) \, ,
\label{dcexp}
\end{equation}
where the $A_j$ are complex amplitudes and the $p_j$ are complex wave 
numbers.
An efficient method of finding the parameters $A_j$ and $p_j$ is described
elsewhere \cite{iectf}.
The evaluation of $I(T)$ is then straightforward:
\begin{equation}
I(T) = - 2 \pi k_B T \, \sum_{j=1}^M \,
\frac{A_j}{{\rm exp} \, (\pi k_B T p_j) - {\rm exp} \, (-\pi k_B T p_j)} \, ,
\label{cexpT}
\end{equation}
which for $T=0$ K gives
\begin{equation}
I(0) = - \sum_{j=1}^M \, \frac{A_j}{p_j} \, .
\label{cexp0}
\end{equation}
The efficiency of the present approach allows to perform calculations
with a large number of ${\bf k}_{\|}$-points in the irreducible part of
the surface Brillouin zone (ISBZ) in order to obtain accurate and reliable
results.
Note also that such calculations have to be done only once and then the
evaluation of the IEC for any reasonable temperature is an easy task.

The effect of finite temperatures on the IEC can be evaluated also
analytically.
The analytical approach assumes the limit of large spacer thickness,
for which all the oscillatory contributions to the energy
integral cancel out with exception of those at the Fermi energy.
The energy integral is then evaluated by a standard saddle-point method
\cite{PBr}.
The general functional form of the temperature-dependence of the interlayer
exchange coupling ${\cal E}_x(T)$ in the limit of a single period is then
given by
\begin{equation} \label{eq_model}
{\cal E}_x(T) = {\cal E}_x(0) \, t(N,T) \, , \;\;\;\;
t(N,T)=\frac{cNT}{\sinh(cNT)} \, .
\end{equation}
Here, $N$ denotes the spacer thickness in monolayers, and $c$ is a
constant which depends on the spacer Fermi surface.
The term ${\cal E}_x(0)$ exhibits a standard $N^{-2}$-dependence
\cite{PBr}, while the scaling factor $t(N,T)$ depends on the product
$N$ and $T$.
In the preasymptotic regime (small spacer thickness) the functional form
of $t(N,T)$ differs from that of~(\ref{eq_model}), particularly in the
case of a complete, but relatively weak confinement due to the rapid
variation of the phase of the integrand which enters the expression for the
IEC \cite{iecta}.
The present numerical technique is free of the above discussed limitations
and can be used to check conclusions of model theories.

\section{Numerical results and discussion}

\subsection{Details of calculations}
Special care has to be devoted to the energy and the Brillouin 
zone integrations. 
For a finite temperature we determine the parameters of the complex 
exponentials in (\ref{dcexp}) through an evaluation of $\Phi(y)$ at 
40 Matsubara energies corresponding to $T=25$ K.
We have verified that the results depend weakly on the actual value of the
parameter $T$.
For $T=0$ K we have tested two energy contours $C$, namely a semicircle 
between the bottom of the band ($E_{min}$) and $E_{F}$, or, alternatively, 
a line contour $E_F+i \, \varepsilon$, $\varepsilon \in (0,\infty)$, 
using a Gaussian quadrature.
The results were very similar in both cases.
Using a line contour avoids possible problems
connected with the phase of a complex logarithm.
Typically a total of 10-15 energy points was used.
A large number of ${\bf k}_{\|}$-points in the ISBZ is needed only for 
energy points close to the real axis, whereby generally a greater 
number is needed for lower temperatures and thicker spacers. 
The number of ${\bf k}_{\|}$-points can significantly be reduced 
for energies well off the real axis.
In particular, for the first energy point on the contour close to
the Fermi energy we typically use 5000-10000 ${\bf k}_{\|}$-points 
in the ISBZ, while for the next 3-4 energy points the number of 
${\bf k}_{\|}$-points is reduced by a factor two for each other point, 
and about 50-100 ${\bf k}_{\|}$-points are taken for all remaining 
energy points on the contour.
The thickness of the spacer, for which well converged results are
obtained, is about 100 spacer layers.

\subsection{Analysis of the results} \label{sec_ares}
The calculated results, namely ${\cal E}_{x}({\theta},N)$, where
$N$ specifies the spacer thickness, can be analyzed in terms of
a discrete Fourier transformation
\begin{equation} \label{eq_ft1}
F(\theta,q) = \frac{1}{p} \, \sum_{N=N_{min}}^{N_{max}}  
N^2 {\cal {E}}_x(\theta,N) \; \exp(i q N) \, ,
\end{equation}
where $p=N_{max} - N_{min}+1$ is the number of values used in the
Fourier analysis, and $N_{min}$ is chosen in order to eliminate 
the effect of very thin spacers, or, to analyze intentionally either
the preasymptotic or the asymptotic region. 
Typically $p$ is about 40.
The background oscillations thus occurring \cite{iectq} are due to 
the discreteness of the Fourier transformation.
The background oscillations can be smoothened using the procedure
described in~\cite{Lan}, namely by multiplying 
$N^{2} {\cal {E}}_x(\theta,N)$ by $C \sin(\pi N/p)/(\pi N/p)$, 
where $C$ is a normalization factor.
The periods of oscillations $\Lambda _{\alpha}$ (in monolayers) are 
then identified with the positions $q_{\alpha}$ of pronounced peaks 
of $\left| F(q_{\alpha})\right| $ as $\Lambda _{\alpha}=2\pi /q_{\alpha}$, 
the amplitudes of oscillations $A_{\alpha}$ are estimated from 
$A_{\alpha}=(2/p)|F(q_{\alpha})|$, and their phases from 
$\phi_{\alpha}=\pi/2 - {\rm Arg} F(q_{\alpha})$, 
($\alpha=1,2, \dots $).
This analysis can be extended to more complicated cases, namely
when the IEC is a function of two variables, e.g.,
as a function of the spacer and cap thicknesses $N$ and $P$, 
respectively.
A two-dimensional discrete Fourier transformation
\begin{equation} \label{eq_ft2}
F_{2} (\theta,q_N,q_P) = 
\sum_{N=N_1}^{N_2} \sum_{P=P_1}^{P_2} 
(N+P)^{2} \, {\cal E}_{2} (\theta,N,P) \,
{\rm e}^{{\rm i}(q_N N+q_P P)} \, 
\end{equation}
is a suitable tool to analyze the quantity 
${\cal E}_{2}(\theta,N,P)$, where the prefactor
$(N+P)^{2}$ is consistent with the asymptotic behavior 
\cite{ieccap1,PBrc} for large spacer and cap thickness.
Strictly speaking, this is quite an obvious choice for the case when the
spacer and cap are formed by the same material, but it can be used
also when the spacer and the cap correspond to different materials
(for more details, see~\cite{ieccap1}).
In (\ref{eq_ft2}) we have introduced the quantity
\begin{equation}
\label{decom3}
{\cal E}_2(\theta,N,P) = {\cal E}_{x}(\theta,N,P) - 
{\cal E}_{0}(\theta,N) \, , \;\;
{\cal E}_{0}(\theta,N)= \lim_{P \rightarrow \infty}
{\cal E}_{x}(\theta,N,P)
\end{equation}
in order to remove a trivial peak in the absolute value of
$F_{2}(\theta,q_N,q_P)$ at $q_N=q_P=0$.
A similar two-dimensional discrete Fourier transformation is also 
useful in the study of the IEC with respect to the thicknesses of 
the spacer and the magnetic slabs.
We note that if one of variables, e.g., the spacer thickness
$N$ is fixed, it is possible to analyze the calculated IEC values 
again with the help of~(\ref{eq_ft1}).

An alternative of calculating the Fourier transform (\ref{eq_ft1})
consists in subdividing the ${\bf k}_{\|}$-integral in~(\ref{eq_iec})
into areas around the critical ${\bf k}_{\|}$-vectors
(callipers) related to the different oscillation periods 
\cite{LNWZD,NLZD}.
In the asymptotic limit each subarea gives then rise to a single oscillation
period, while in the preasymptotic regime the resulting division into
different periods is only qualitatively valid.
In a sense this method bridges the present method of discrete Fourier
transformations and the purely asymptotic treatment of calculating only
the behavior of the critical ${\bf k}_{\|}$-vectors 
(see Section~\ref{sec_asex}).

\subsection{Asymptotic expansion}\label{sec_asex}
Model studies \cite{PBr,JMa} indicate that in the 
asymptotic region, i.e., for large spacer and magnetic layer thickness,
and for a random sample, the general form of the spacer-thickness 
dependence of the IEC is given by
\begin{equation} \label{eq_asympt}
{\cal E}_{x} =  {\rm Im} \, \sum_{\alpha} \frac{Z_{\alpha}}{N^{2}} \;
\exp(i Q_{\alpha} N) \, .
\end{equation}
Here the sum runs over all possible periods $\alpha$, the quantities
$Z_{\alpha}$ and $Q_{\alpha}$ are the complex amplitudes and complex 
stationary points (callipers), respectively, defined in the following 
manner 
\begin{equation}
Z_{\alpha}=A_{\alpha} \, \exp(i \Phi_{\alpha}) \, , \;\;\;\;\;\;\;\;
Q_{\alpha} = q_{\alpha}+ i \lambda_{\alpha} \, .
\end{equation}
The quantities $A_{\alpha}$ and $\Phi_{\alpha}$ are the amplitudes and
phases of coupling oscillations, $p_{\alpha}=2 \pi /q_{\alpha}$ their
periods, and the quantity $\lambda_{\alpha}$ characterizes the damping 
of oscillations due to the effect of alloying in the sample determined
at the Fermi energy.
In the limit of non-random samples, $\lambda_{\alpha}=0$.

The parameters in (\ref{eq_asympt}) can be extracted from a detailed
knowledge of the spacer Fermi surface \cite{St}.
We briefly sketch a numerical way of determining of the parameters
of this asymptotic expansion which requires the knowledge of the
integrand of (\ref{eq_iecpi}) for a set of ${\bf k}_{\|}$-points in
the neighborhood of the stationary points ${\bf k}_{\|}^{(\alpha)}$.

The expression (\ref{eq_iecpi}) for IEC at $T=0$~K can be rewritten as
\begin{equation} \label{eq_iecpi2}
{\cal E}_{x} = \frac{1}{N_{\|}} \, {\rm Im} \, \sum_{{\bf k}_{\|}} \, 
Y({\bf k}_{\|}) \, , \;\:
Y({\bf k}_{\|}) = \frac{1}{\pi} \, \int_{C} \, f(z,0) 
\; {\rm tr}_{L} \, \ln {\cal M}({\bf k}_{\|}, z) \, {\rm d} z \, .
\end{equation}
The integration with respect to the energy variable 
is performed numerically.
The function $Y({\bf k}_{\|})$ for large $N$ decreases as $O(1/N)$
and behaves like
\begin{equation} \label{phiasympt}
Y({\bf k}_{\|}) =
\frac{g({\bf k}_{\|})}{N} 
\, {\rm exp}(iN\phi({\bf k}_{\|})) \, ,
\end{equation}
where the pre-exponential factor $g({\bf k}_{\|})$ is a smooth 
function of ${\bf k}_{\|}$ and the phase $\phi({\bf k}_{\|})$
has one, or more stationary points in the SBZ that correspond 
to callipers of the spacer Fermi surface such that
$\nabla_{{\bf k}_{\|}} \phi({\bf k}_{\|})=0$.
The integral over the SBZ in (\ref{eq_iecpi2})
can be evaluated using the stationary-phase method.
The contribution of a stationary point 
${\bf k}_{\|}^{(\alpha)} \equiv (k_x^{(\alpha)},k_y^{(\alpha)})$
is found in the following way:
the integration limits are extended to infinity, and the phase 
function $\phi({\bf k}_{\|})$ is approximated by a quadratic 
function of ${\bf k}_{\|} \equiv (k_x,k_y)$
in the vicinity of the stationary point,
\begin{eqnarray} \label{quadrapr}
\phi({\bf k}_{\|}) &=& \phi({\bf k}_{\|}^{(\alpha)})+
\sum_{i,j = x,y} Q_{ij} 
(k_{i}-k_{i}^{(\alpha)})(k_{j}-k_{j}^{(\alpha)}) \nonumber \\
&=& \sum_{i,j = x,y} Q_{ij} k_{i} k_{j}
+\sum_{i = x,y} P_i k_i +
\phi({\bf k}_{\|}^{(\alpha)}) \, .
\end{eqnarray}
The expansion coefficients $Q_{ij}$, $P_{i}$,
and $\phi({\bf k}_{\|}^{(\alpha)})$ are determined 
by a least-square fit to values of $\phi({\bf k}_{\|})$ 
calculated in the vicinity of ${\bf k}_{\|}^{(\alpha)}$.
This procedure allows to eliminate numerical inaccuracies with
respect to both the values of $Q_{ij}$ and the position of the 
stationary point ${\bf k}_{\|}^{(\alpha)}$, and it is applicable 
even for disordered surfaces.
By inserting (\ref{phiasympt}) and (\ref{quadrapr}) 
into~(\ref{eq_iecpi2}) we find
\begin{eqnarray} \label{exasympt2}
{\cal E}_{x} &\approx& 
\frac{1}{\pi N V_{\rm SBZ}} \, {\rm Im} \, \Biggl\{
g({\bf k}_{\|}^{(\alpha)}) \times \nonumber \\
& & \int\!\!\!\int_{D} 
{\rm exp} \Bigl[i N \Bigl(\phi({\bf k}_{\|}^{(\alpha)}) + 
\sum_{i,j = x,y} Q_{ij} (k_{i}-k_{j}^{(\alpha)})(k_{j}-k_{j}^{(\alpha)})
\Bigr) \Bigr] {\rm d} k_x \, {\rm d} k_y \Biggr\} \nonumber \\
& = & \frac{\pi}{N^2 V_{\rm SBZ}} \, {\rm Im} \, \Biggl\{ 
\frac{g({\bf k}_{\|}^{(\alpha)})} {\sqrt{-\det|Q|}} \,
{\rm exp}\Bigl[i N \phi({\bf k}_{\|}^{(\alpha)})\Bigr]
\Biggr\} \, ,
\end{eqnarray}
where the two-dimensional integration region $D$ extends to infinity,
and $V_{\rm SBZ}$ denotes the volume of the SBZ.
The second line in (\ref{exasympt2}) is obtained by diagonalizing
the quadratic form in the exponent (\ref{quadrapr}) and by evaluating 
the resulting one-dimensional Gaussian-like integrals. 
The identification of the parameters is now straightforward, namely
\begin{equation} \label{identif}
Z_{\alpha} = \frac{\pi}{V_{\rm SBZ}} \, 
\frac{g({\bf k}_{\|}^{(\alpha)})} {\sqrt{-\det|Q|}}
\, , \quad 
Q_{\alpha} = \phi({\bf k}_{\|}^{(\alpha)})
\, .
\end{equation}

\subsection{Free-electron limit}

The numerical efficiency of the present formalism offers an interesting
possibility of testing model theories \cite{PBr}.
The simplest of such models is the free-electron model,
because of a spherical Fermi surface with a single critical vector at
${\bf k_{\|}}=0$ and a trivial correspondence between the value of
the oscillation period and the band-filling.
The free-electron model can be easily simulated by the present formalism
by replacing the true metallic potentials by flat potentials (the 
empty-sphere model).
In this case the potential functions~(\ref{pfnc}) are
analytical functions of the lattice constant.
For a suitable choice of the lattice constant and the position of the
Fermi energy it is irrelevant what lattice and layer stacking is used,
e.g., the fcc(001)-stack is the simplest choice.
On the other hand such a model is free of the limitations usually adopted
\cite{PBr}, e.g., the assumption of large spacer and magnetic slabs 
thicknesses, or the approximate evaluation of the energy integral for
the case of finite temperatures.

\subsection{Numerical illustrations}

In Fig.~1 $N^{2} {\cal E}_{x}(N)$ is displayed as a function of the 
spacer thickness $N$ for two semi-infinite Co(001) subsystems sandwiching 
an fcc-Cu spacer.
The corresponding discrete Fourier transformation in Fig.~2
shows a pronounced short-period oscillations of 2.53 monolayers (MLs)
while the long-period oscillations are suppressed in this geometry
\cite{iecsgf,iectq,NLZD}.
The results are insensitive to the choice of the lower and upper
index in the summation in (\ref{eq_ft1}) provided the preasymptotic
region is excluded \cite{iectq}.
 
\vskip -10mm 
\begin{figure}[h]
\includegraphics[width=.5\textwidth]{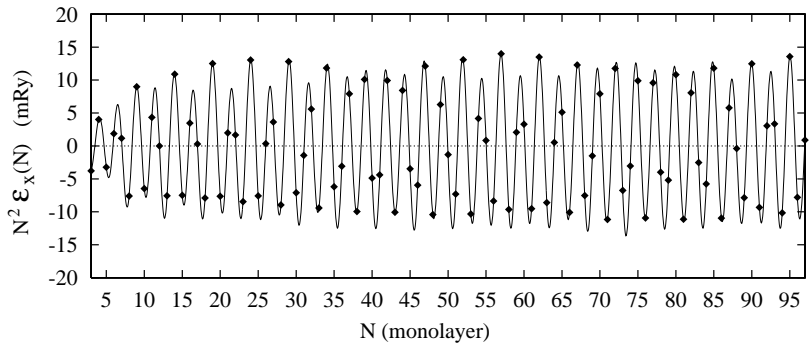}
\vskip -110mm 
\caption[]{Exchange coupling $N^2 {\cal E}_x(N)$ at $T=0$~K as a function
of the spacer thickness $N$ for two semi-infinite fcc Co(001) subsystems
sandwiching a Cu spacer. Diamonds refer to the calculated values,
the full line (back Fourier transform) serves as a guide to the eye}
\label{Fig.1}
\end{figure}

For a large enough $N$ the IEC can be approximated  by the asymptotic
form in~(\ref{eq_asympt}).
The amplitude, phase, and the wave-vector entering this expression can
be determined from the calculated ${\cal E}_{x}(N)$ in the manner as described
in Sec.~\ref{sec_ares} and the asymptotic result (\ref{eq_asympt}) was 
compared with the calculated results for a large set of systems including
both ideal and alloyed semi-infinite fcc(001) magnetic subsystems
sandwiching a Cu-spacer: overall good agreement was found \cite{iecams}. 
An example of the complex amplitude for this case is presented in Fig.~3 
illustrating the insensitivity of the phase to elements which form
the magnetic layers.
It is seen that phases corresponding to Co,

\begin{figure}[h]
\sidecaption
\includegraphics[width=.5\textwidth]{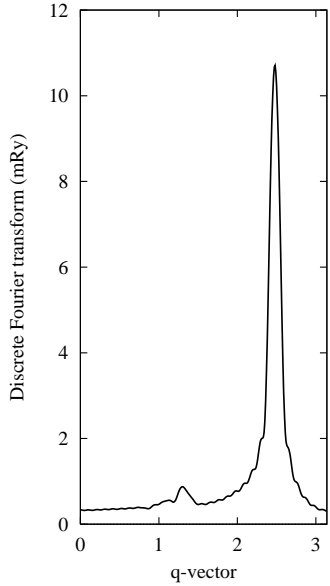}
\caption[]{Absolute value of the discrete Fourier transformation
of $N^2 {\cal E}_x(N)$ for a finite set of spacer layers ($N$=20--80)
corresponding to two semi-infinite fcc Co(001) subsystems sandwiching
a Cu spacer. The temperature is $T=0$~K }
\label{Fig.2}
\end{figure}

\noindent Fe$_{50}$Ni$_{50}$, and 
Fe$_{1/3}$Ni$_{1/3}$Co$_{1/3}$ which have the same average electron numbers
$N_{el}$=9 are nearly the same \cite{iecams}.

\begin{figure}[h]
\sidecaption
\includegraphics[width=.5\textwidth]{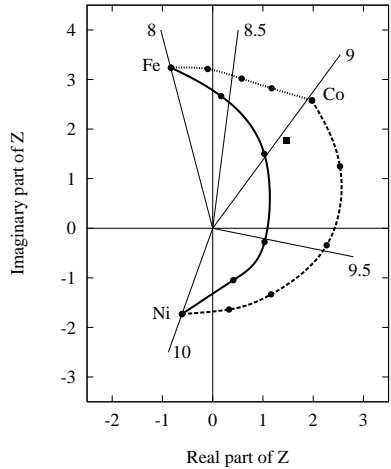}
\caption[]{Complex amplitude $Z^{1/2}=A^{1/2} \, {\rm e}^{i \Phi/2}$,
where $A$ and $\Phi$ are the oscillation amplitude and phase,
respectively, for a semi-infinite fcc(001) subsystems
formed by Fe, Co, Ni, their binary alloys (bullets), and the ternary
alloy Fe$_{1/3}$Co$_{1/3}$Ni$_{1/3}$
(square) sandwiching a Cu spacer. The units are (mRy)$^{1/2}$.
The dotted, dashed, and full lines connect various alloys and serve
as a guide to the eye.  The rays starting at the origin show
approximately the phase corresponding to the indicated average
number of valence electrons }
\label{Fig.3}
\end{figure}

The IEC depends on the temperature $T$ via a factor 
${x/{\rm sinh}(x)}$, ${x}=c NT$, where $T$ is the temperature and $N$
the spacer thickness.
This remarkable result of model theories \cite{PBr} was verified by
calculations such as illustrated in Fig.~4.

\pagebreak

\vskip 10mm
\begin{figure}[h]
\sidecaption
\includegraphics[width=.5\textwidth]{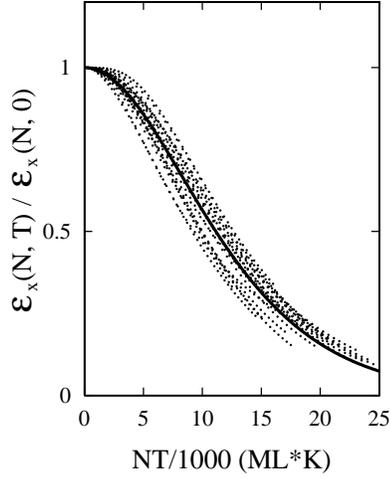}
\caption[]{${\cal E}_{x}(N,T)/{\cal E}_{x} (N,T=0)$ plotted as a function
of $\zeta=NT$ for a trilayer consisting of semi-infinite fcc Co(001)-slabs
sandwiching a Cu spacer.  The thick line refers to
${\rm x/sinh(x)}$, ${\rm x}= c \, NT$ with $c=0.000195$ obtained by
a least square-fit to the computed data}
\label{Fig.4}
\end{figure}
\vskip 5mm

\noindent The IEC depends in an oscillatory manner not only on the spacer 
thickness $N$ but as well on the thickness $P$ of a covering cap.
The oscillations are around a biased value which corresponds to
coupling for a given spacer thickness assuming a semi-infinite cap.
This phenomenon is illustrated in Fig.~5 in terms of discrete Fourier
transformations with respect to the spacer and the

\vskip -25mm
\begin{figure}[h]
\includegraphics[width=.75\textwidth]{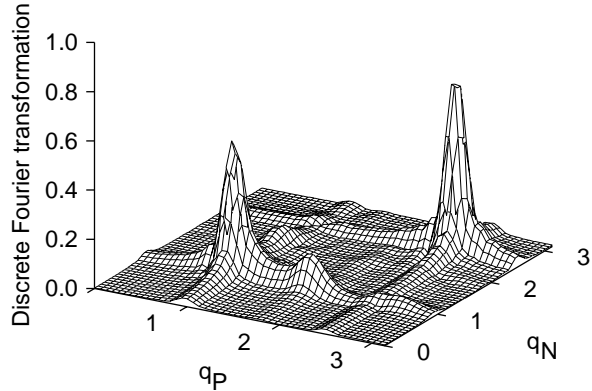}
\caption[]{Absolute values of the discrete two-dimensional Fourier
transformation of $(N+P)^{2} \, {\cal E}_2(N,P)$  with respect to
the spacer and the cap thickness in the case of two magnetic slabs
each five monolayers thick with a Cu-substrate, a Cu-spacer, and
a Cu-cap. For a definition of ${\cal E}_2(N,P)$ see the text}
\label{Fig.5}
\end{figure}

\noindent cap thickness (see 
Sec.~\ref{sec_ares}) for a sample consisting of a semi-infinite
fcc-Cu(001) substrate, left and right magnetic layers each five MLs thick,
a spacer with varying thickness $N$, and a Cu-cap of varying thickness $P$.
Fig.~5 shows: (i) long-period oscillations (missing in Fig.~2) in addition 
to the short-period ones, and (ii) oscillations with respect to the cap 
thickness which are exactly the same as for the spacer because both are 
controlled by the same Fermi surface, namely that of fcc-Cu.
The more complicated case of different spacer and cap materials is
discussed in~\cite{ieccap1,ieccap2}. 

\vskip 5mm
\begin{figure}[h]
\sidecaption
\includegraphics[width=.5\textwidth]{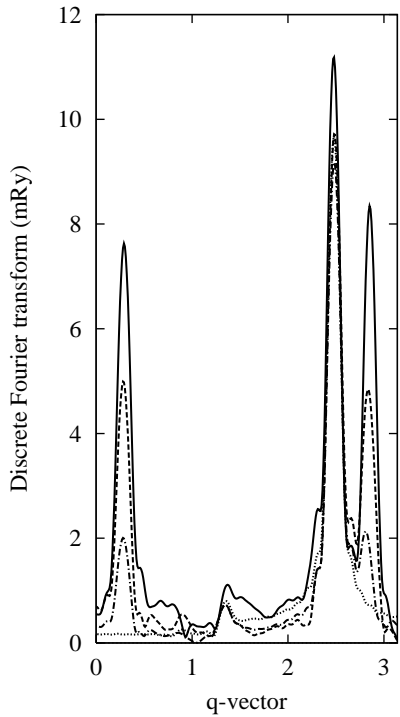}
\caption[]{ Absolute values of the discrete Fourier transformation of
$N^{2} \, {\cal E}_x(N)$ for two semi-infinite fcc Co$_{50}$Fe$_{50}$(001)
subsystems sandwiching a Cu spacer with different kinds of chemical order
in magnetic layers: ({\bf a}) $S=1$ (complete $c(2 \times 2)$-order,
full line), ({\bf b}) $S=0.8$ (dashed line), ({\bf c}) $S=0.5$
(dashed-dotted line), and ({\bf d}) $S=0.0$ (disordered case,
dotted line). The temperature is $T=0$~K }
\label{Fig.6}
\end{figure}
\vskip 10mm

Ordering in the spacer \cite{iecsor} or in the magnetic layers
\cite{iecmor} can induce new periods due to the formation of 
two-dimensional sublattices.
The situation is particularly interesting for a 
$c(2 \times 2)$-ordering in magnetic layers sandwiching an ideal 
Cu-spacer \cite{iecmor}.
As illustrated in Fig.~6 for full ordering two new periods with 
complementary periods and phases are formed in addition to 
a conventional short-period due to a fcc-Cu spacer \cite{iecmor}.
These new periods vanish in the completely disordered case.

Finally, the effect of disorder in the spacer \cite{iecas} is 
illustrated in Fig.~7.
Alloying of Cu with Ni decreases the number of average valence electrons
and leads to a contraction of the alloy Fermi surface, and in turn
to a reduction of the coupling oscillations.
The opposite behavior has to be expected for alloying of Cu with Zn,
whereas only a small concentration dependence of the periods for 
the CuAu case is seen.
The amplitudes of the oscillations are generally reduced by alloying, 
and in the case of CuZn spacer they are even
\newpage

\vskip -20mm
\begin{figure}[h]
\includegraphics[width=.75\textwidth]{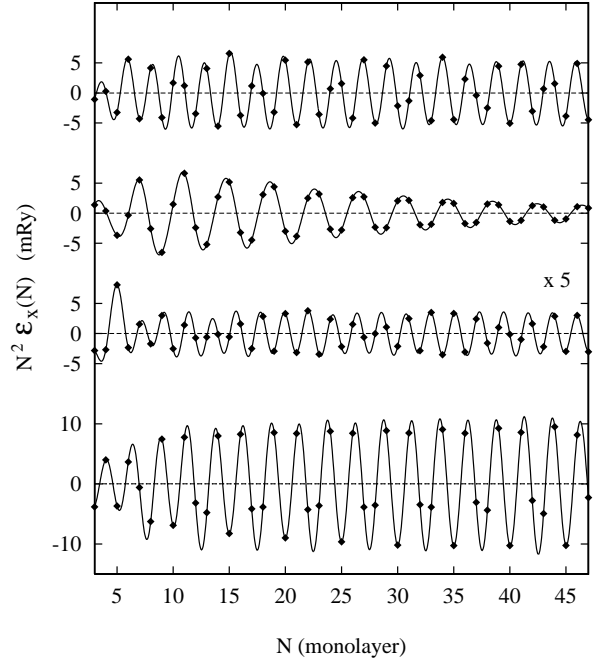}
\vskip 15mm
\caption[]{Exchange coupling $N^2 {\cal E}_x(N)$ at $T=0$~K as a function
of the spacer thickness $N$ for two semi-infinite fcc Co(001) subsystems
sandwiching a spacer of (from bottom to top) Cu, Cu$_{75}$Ni$_{25}$
(multiplied by a factor 5), Cu$_{50}$Zn$_{50}$, and Cu$_{50}$Au$_{50}$.
Diamonds refer to the calculated values, the full line (back Fourier
transform) serves as a guide to the eye}
\label{Fig.7}
\end{figure}

\noindent exponentially damped.
The different behavior of the amplitudes can be related to differently
large disorder in the neighborhood of relevant extremal points
of the alloy Fermi surfaces. 

\subsection{List of published applications}

We briefly review applications of the formalism developed in previous 
sections to specific problems.
Additional details concerning formalism and not discussed here in
details, e.g., the expansion of the IEC expression in terms of the small
parameter $1-\cos(\theta)$ or the details concerning the numerical
verification of the vertex-cancellation theorem, can be found 
in~\cite{iectq,iecvtx}, respectively.
The influence of surface roughness (fluctuating spacer thickness and
diffusion at the interface between spacer and magnetic layers)
on the oscillation amplitudes was studied in~\cite{iecdb}.
The effect of alloying in the spacer \cite{iecas} on the oscillation periods
and their amplitudes, and in magnetic layers \cite{iecams} on the oscillation
amplitudes and phases was also studied in detail for the trilayer
system Co/Cu/Co(001).
Ordering in disordered spacers \cite{iecsor} and/or magnetic \cite{iecmor}
layers lead to a formation of new periods not present in ideal spacers.
Oscillations of the IEC can originate not only due to the spacer
but also from adlayers or cap layers.
We refer the reader interested in this problem to a recent detailed
study \cite{ieccap1,ieccap2}.
Finally, the study of the temperature dependence of the IEC and of the
combined effect of the temperature and disorder is subject of
very recent papers \cite{iectf,iectfa}, respectively.

\section{Conclusions}

We have derived closed expressions for the exchange coupling 
between two magnetic subsystems separated by a non-magnetic spacer
with a relative angle $\theta$ between the orientations of the
magnetizations in the magnetic slabs.
The derivation is based on a surface Green function formalism.
The numerical effort scales linearly with the thickness 
of both the spacer and the magnetic slabs.
The formulation allows also for an efficient evaluation of the 
temperature dependence of the coupling amplitudes.
Numerical examples were chosen to illustrate the theoretical
aspects rather than to give a comprehensive overview
of results obtained by the present formalism or by related methods.

We wish now briefly to mention some unsolved problems.
The following list is neither complete nor are the problems
listed according to their importance:
(i) The oscillatory dependence of the IEC on the thickness of the 
magnetic slabs was not yet systematically investigated on an 
{\it ab initio} level.
Existing calculations \cite{iectq,KKP,NLZD} were performed for too 
thin magnetic slabs to relate occurring oscillations to extremal 
points of spin-polarized Fermi surfaces;
(ii) The problem of biquadratic and higher order terms also did not 
receive a proper attention on an {\it ab initio} level.
A relevant problem is a systematic study of situations for which the
non-collinear (biquadratic) coupling can dominate.
Obviously, it can happen most probably for the spacer thicknesses
for which the IEC values are close to the transition between the
F and AF couplings \cite{gmr}.
In addition, it remains to be seen whether a theoretical description
of biquadratic coupling has to be based on a fully relativistic
spin-polarized level;
(iii) The study of superstructures in the spacer and/or in the magnetic
slabs (see Sec.~\ref{sc_ext}) offers a possibility of a deeper
insight into the physical nature of the IEC because of new periods, 
which are connected with the extremal vectors of the spacer material 
in a more sophisticated manner than in the canonical cases of Cu or Cr 
spacers;
(iv) The study of oscillatory behavior of exchange interaction across
magnetic spacers is of great interest.
One possibility here is to employ the method of infinitesimal
rotations \cite{mft2,iectq};
(v) The study of exchange coupling through the semiconducting or, 
more generally, through an insulating spacer where one expects 
exponential rather than $N^{-2}$-decay has remained limited until now 
to model studies \cite{PBr};
(vi) The study of alloying in the spacer, magnetic layers and at 
interfaces has to be extended to new interesting systems. 
It offers a straightforward method to obtain valuable informations 
concerning alloy Fermi surfaces, in particular for the case of
alloyed spacers; 
and, finally 
(vii) The study of the IEC through spacers with complex Fermi surfaces,
in particular through the transition metal spacers.\\

\noindent {\it Acknowledgements} This work is a part of activities 
of the Center for Computational Material Science sponsored by the 
Academy of Sciences of the Czech Republic.
Financial support for this work 
was provided by the Grant Agency of the Czech Republic 
(Project No. 202/97/0598), the Grant Agency
of the Academy Sciences of the Czech Republic (Project A1010829), 
the Center for the Computational Materials Science
in Vienna (GZ 45.442 and GZ 45.420), and the TMR Network 'Interface
Magnetism' of the European Commission (Contract No. EMRX-CT96-0089).

\appendix

\section{Vertex cancellation theorem}

We present here a general discussion of exchange interactions in the presence 
of substitutional disorder. The results given here are used in the present 
paper to study interlayer exchange interactions, but they are also applicable 
for studying exchange interactions within a ferromagnet, exchange 
stiffnesses, spin-wave energies, etc. The principal result is the 
``vertex cancellation theorem'' of Bruno {\em et 
al.\/} \cite{iecvtx}. In here we give an alternative, more general, 
derivation of this result.

Let $\bu \equiv \{\bu_\bR\}$ be a particular 
configuration of the local moments, where $\bu_\bR$ is
a unit vector pointing in the direction of the $\bR$-th 
local moment.
We are interested in the variation of the thermodynamic 
grandcanonical potential
\begin{equation}
\Omega_\bu = 
- \, \frac{1}{\pi} \mbox{ Im} \int_{-\infty}^{+\infty} \!\!  \,
 f(E,T) \, \mbox{Tr} \left< \ln g_\bu (E+i0^+) \right> {\rm d}E 
\label{appa1}
\end{equation}
with respect to $\bu$.
The Green function $g_\bu (z)$ for a particular alloy
configuration is defined from
the potential function $P_\bu (z)$ corresponding to $\bu$ as
\begin{equation}
g_\bu (z) = \left( P_\bu (z) - S \right)^{-1} \ .
\label{appa2}
\end{equation}
An immediate consequence of (\ref{appa2}) is a trivial
commutator relation to be used below, namely
\begin{equation}
\left[ P_\bu (z) ; g_\bu (z) \right]_{-} = 
\left[ S ; g_\bu (z) \right]_{-}  \ ,
\label{appa3}
\end{equation}
where $\left[ A ; B \right]_{-} \equiv AB-BA$.
The configuration averaged Green function 
$\left< g_\bu (z) \right> \equiv
\overline{g}_\bu (z)$ is usually formulated in terms of
the coherent potential function
${\cal P}_\bu (z)$ as
\begin{equation}
\overline{g}_\bu (z) = 
\left( {\cal P}_\bu (z) - S \right)^{-1} \ ,
\label{appa4}
\end{equation}
which leads to a relation analogous to (\ref{appa3}),
\begin{equation}
\left[ {\cal P}_\bu (z) ; \overline{g}_\bu (z) \right]_{-} 
 = \left[ S ; \overline{g}_\bu (z) \right]_{-} \ . 
\label{appa5}
\end{equation}
In general, the averaging in (\ref{appa1}) cannot be reduced 
to $\ln \, \overline{g}_\bu (z)$ and an evaluation of the 
so-called vertex corrections is necessary. 
We shall show, however, that the variation of (\ref{appa1})
due to an infinitesimal change of $\bu$ takes a simple form.

Let us consider the variation of the potential functions 
$P_\bu (z)$ in more detail. 
To each lattice site $\bR$ we associate a non-random vector 
$\bTheta_\bR \equiv \theta_\bR \, \hat{\bf n}_\bR$, where
$\hat{\bf n}_\bR$
refers to the axis of rotation and $\theta_\bR$ to rotation angle by
which the reference orientation $\hat{\bf u}_{0,\bR}$ is transformed
into $\hat{\bf u}_{\bR}$.
The transformed potential functions are therefore given by the
following similarity transformation
\begin{equation}
P_\bu (z) = U_\bTheta  \, 
P_{\bu_0} (z) \, U_\bTheta^{-1} \ ,
\label{appa_pudef}
\end{equation}
where the rotation matrix $U_\bTheta$ in (\ref{appa_pudef}) is defined as
\begin{eqnarray}
\left( U_\bTheta \right)_{\bR L s ,\bR ' L' s' } 
 & = & \delta_{\bR ,\bR'} \, \delta_{L,L'} \times 
                                             \nonumber\\
 && \left[ \cos \left(\frac{\theta_\bR}{2}\right)\, 
{\mbox{\small  l} \! \! \hspace{0.017cm} 1} \, - \, i \, 
\sin\left( \frac{\theta_\bR}{2}\right) \, \hat{\bf n}_{\bR} 
\cdot \bsigma \right]_{s,s'} \, .
\label{appa_udef}
\end{eqnarray}
The symbol $\bsigma$ in (\ref{appa_udef}) denotes the vector of
the standard $2 \times 2$ Pauli matrices and 
${\mbox{\small  l} \! \! \hspace{0.017cm} 1}$ is the $2 \times 2$
unit matrix.
The first-order change of $P_{\bu, \bR} (z)$ caused by an additional
infinitesimal rotation $\delta {\bf v}_\bR$ is then expressed as
\begin{equation}
\delta P_\bu (z) = 
\left[ \delta K ; P_\bu (z) \right]_{-} \ ,
\label{appa6}
\end{equation}
where the matrix elements of the operator 
$\delta K = U_{\delta {\bf v}} - 1$ are explicitly given by
\begin{equation}
\left( \delta K \right)_{ \bR L s , \bR ' L' s' }  
= \delta_{\bR ,\bR'} \, \delta_{L,L'} \, 
\frac{(-i)}{2} \,
\left[ \bsigma \cdot \delta {\bf v}_\bR \right]_{s,s'} \, .   
\label{appa7}
\end{equation}
The introduced infinitesimal rotation vectors
$\delta {\bf v}_\bR$ satisfy 
$U_{\delta {\bf v}} \, U_{\bTheta} = U_{\bTheta + \delta \bTheta}$
whereas, in general, 
$U_{\delta \bTheta} \, U_{\bTheta} \neq U_{\bTheta + \delta \bTheta}$.
Let us note that $\delta K$ is a non-random site-diagonal
operator.

The first-order variation of $\mbox{Tr} \left< \ln g_\bu (z) \right>$
can be now formulated using (\ref{deriva}, \ref{appa6}) as
\begin{equation}
\delta \mbox{Tr} \left< \ln g_\bu (z) \right> = 
 - \, \mbox{Tr} \left< g_\bu (z) \,  
\left[ \delta K ; P_\bu (z) \right]_{-} \right> \ , 
\label{appa8}
\end{equation}
which can be rewritten by applying the permutation invariance 
of the trace and (\ref{appa3}, \ref{appa5}) as
\begin{eqnarray}
\delta \mbox{Tr} \left< \ln g_\bu (z) \right> 
& = & - \, \mbox{Tr} \left\{ \delta K \,
\left< \left[ P_\bu (z) ;
g_\bu (z) \right]_{-} \right> \right\}
                           \nonumber \\
& = & - \, \mbox{Tr} \left\{ \delta K \,
\left< \left[ S ;
g_\bu (z) \right]_{-} \right> \right\}
                           \nonumber \\
& = & - \, \mbox{Tr} \left\{ \delta K \,
\left[ S ;
\overline{g}_\bu (z) \right]_{-}  \right\}
                           \nonumber \\
& = & - \, \mbox{Tr} \left\{ \delta K \,
\left[ {\cal P}_\bu (z) ;
\overline{g}_\bu (z) \right]_{-} \right\} \ .
\label{appa9}
\end{eqnarray}
By using the permutation invariance of the trace
once again, (\ref{appa9}) can be given the final form
\begin{equation}
\delta \mbox{Tr} \left< \ln g_\bu (z) \right> =  
- \, \mbox{Tr} \left\{ \overline{g}_\bu (z) \,  
\left[ \delta K ; {\cal P}_\bu (z) 
\right]_{-} \right\} \ .
\label{appa10}
\end{equation}
Let us note that (\ref{appa10}) was derived in a formally
exact alloy theory, but is valid in the CPA as well. 
Within the CPA, the result (\ref{appa10}) has an obvious
interpretation: the r.h.s. describes the variation of 
$\mbox{Tr} \ln \overline{g}_\bu (z)$ induced by performing on 
the site-diagonal coherent potential functions 
${\cal P}_{\bu , \bR} (z)$ the same rotations (\ref{appa7}) as 
applied to the potential functions $P_{\bu , \bR} (z)$; note however, 
that this is {\em not\/} equal to the infinitesimal change of the 
true {\em self-consistent\/} CPA coherent potential function.

Thus, the torque acting on the moment at site $\bR$ due to the exchange 
interactions is given by
\begin{eqnarray}
{\bf\Gamma}_{\bu ,\bR} &\equiv & 
-\, \frac{\delta \Omega_{\bu}}{\delta {\bf v}_{\bR} }
= -\, \frac{1}{\pi} \int_{-\infty}^{+\infty} f(E,T)\times
\nonumber \\ && \mbox{ Im Tr}
\left\{ \overline{g}_{\bu}(E+i0^+) \,
\frac{(-i)}{2} \, \left[ {\sf\Pi}_\bR \, \bsigma \, ; \, 
{\cal P}_{\bu}(E+i0^+) \right]_{-}
\right\}  {\rm d}E  \ ,
\label{eq_torque}
\end{eqnarray}
where ${\sf\Pi}_\bR$ is a projector on site $\bR$.
This exact result constitutes the ``vertex cancellation theorem'' for the 
torque. Its usefulness arises from the fact that the ``vertex 
corrections'' have been eliminated. 

In order to compute the difference of thermodynamic grandcanonical 
potential between 
two local moment configurations $\bu_1$ and $\bu_2$ in the CPA, 
we use a theorem due to
Ducastelle \cite{FD}, which states that the 
thermodynamic grandcanonical potential,
considered as a functional $\tilde{\Omega}[{\cal P}, P]$ of the 
independent variables ${\cal P}$ and $P$, is stationary 
with respect to ${\cal P}$ when the latter satisfies the CPA 
self-consistency condition. This means that a 
first-order error in ${\cal P}_\bu$ gives only a second-order 
error in $\Omega_\bu$. 
Let us approximate ${\cal P}_\bu (z)$ by
\begin{equation}
{\cal P}_\bu (z) \approx
{\cal P}^\prime_\bu (z) \equiv U_\bTheta  \, 
{\cal P}_{\bu_0} (z) \, U_\bTheta^{-1} \ ,
\end{equation}
i.e., we assume that ${\cal P}_\bu (z)$ is transformed 
like $P_\bu(z)$ under a rotation 
of the local moment direction. This can be expected to be 
a good approximation, provided the condition 
\begin{equation}
m_\bR \, \left| \frac{{\rm d}\bTheta_\bR}{{\rm d}\bR} \right| 
\ll k_F \, q_\bR 
\label{eq_cond}
\end{equation}
is satisfied, 
where $q_\bR$ and $m_\bR$ are respectively the charge and 
spin moment at site $\bR$. 
We then get 
\begin{equation}
\overline{g}_\bu (z) \approx 
\overline{g}_\bu^{\,\prime} (z) 
\equiv \left( {\cal P}_\bu^\prime (z) - S \right)^{-1} .
\label{appa_gpdef}
\end{equation}
Replacing ${\cal P}_\bu$ by 
${\cal P}_\bu^\prime$ and $\overline{g}_\bu$ by 
$\overline{g}_\bu^{\,\prime}$ in~(\ref{appa10}),
we obtain
\begin{equation}
\delta \mbox{Tr} \left< \ln g_\bu (z) \right> \approx  
\delta \mbox{Tr} \, \ln \overline{g}_\bu^{\,\prime} (z) \ ,
\end{equation}
and integrating over the angles, we get
\begin{eqnarray}
{\Omega}_{\bu_1} - {\Omega}_{\bu_2} \approx && -\, \frac{1}{\pi} 
\int_{-\infty}^{+\infty} f(E,T) \times \nonumber \\
&& \mbox{ Im Tr} \left[ 
\ln\overline{g}^{\,\prime}_{\bu_1} (E+{\rm i}0^+) 
-\ln\overline{g}^{\,\prime}_{\bu_2} (E+{\rm i}0^+) 
\right]  {\rm d}E  \ ,
\label{eq_vtx2}
\end{eqnarray}
which constitutes the ``vertex cancellation theorem'' for exchange energies. 
Note that we have derived here a form of the ``vertex cancellation 
theorem'' within the CPA since this is the scheme which is used in practical 
calculations; however, one can prove that the same result holds if one takes 
the exact solution to the configuration averaging problem.

In the case of interlayer coupling, the condition (\ref{eq_cond}) is
satisfied even for large rotation angles, 
because ${{\rm d}\bTheta_\bR}/{{\rm d}\bR}$ differs from zero only 
in a region where $m_\bR$ is negligible. This was confirmed by explicit 
numerical calculations in~\cite{iecvtx}.

\section{The interface-interface part of the grandcanonical potential}

In this Appendix we derive the basic relations for an evaluation of 
the IEC within the interface-interface interaction formulation.

The subsystems ${\cal L}$ and ${\cal R}$ can be downfolded using the 
formula (\ref{parttl})
\begin{eqnarray}
& & {\rm Tr} \, {\rm ln}\, (P-S) = 
 {\rm Tr}_{\cal L}\,{\rm ln}\, \Bigl[ P-S \Bigr]
+{\rm Tr}_{\cal R}\,{\rm ln}\, \Bigl[ P-S \Bigr]
\nonumber \\
&+&{\rm Tr}_{\cal C}\,{\rm ln}\, \Bigl[ (P-S)_{\cal CC}-
(P-S)_{\cal CL}\frac{{\cal L}}{P-S}(P-S)_{\cal LC} \nonumber \\
&-& (P-S)_{\cal CR}\frac{{\cal R}}{P-S}(P-S)_{\cal RC} \Bigr] \, .
\label{parttrlog}
\end{eqnarray}
The first two terms are independent of the rotation angle $\theta$
and, consequently, they do not contribute to the exchange energy
${\cal E}_{x}({\theta})$.
We are thus left with a quantity which is limited to the subspace 
${\cal C}$ only.
It is now easy to identify the individual terms in (\ref{parttrlog}).
The potential function blocks between different subspaces such 
as $P_{\cal LC}$ or $P_{\cal CR}$ are zero because the potential
function $P$ is site-diagonal. 
The blocks of $S$ between neighboring subspaces do not vanish,
but the non-zero subblocks connect only neighboring 
principal layers.
The important part of the ${\rm Tr\, ln}\, (P-S)$ is then reduced to
\begin{equation}
{\rm Tr}_{\cal C}\,{\rm ln}\, (P-S)_{\cal CC}+
{\rm Tr}_{\cal C}\,{\rm ln}\, \Bigl[ 1 
-\frac{\cal C}{P-S} S_{10}{\cal G}_{\cal L}S_{01}
-\frac{\cal C}{P-S} S_{01}{\cal G}_{\cal R}S_{10} 
\Bigr] \, .
\label{important}
\end{equation}
The first term is independent of $\theta$ and thus does not 
contribute to the exchange energy.
The second term can be simplified using the two-potential formula
(\ref{twopotlog}).
We identify $G^{(0)}={\cal C}/(P-S)$,
$v_1=S_{10}{\cal G}_{\cal L}S_{01}$, and 
$v_2=S_{01}{\cal G}_{\cal R}S_{10}$. 
The t-matrices are then identical with the $\tau$-matrices, and 
the potentials $v_1$ and $v_2$ are equal to the embedding potentials
$\Gamma_1$ and $\Gamma_2$.
In this way we find the expression for the grandcanonical potential
\begin{eqnarray}
& &\Omega(\theta,T,\mu) = 
\Omega_0(T,\mu) \nonumber \\
& &- \frac{1}{\pi}\:{\rm Im} \int_{-\infty}^{\infty}
\,  f(E,T,\mu) \,{\rm Tr_1\: ln}\: 
\Bigl[
1 - g_{1N}(z) \tau_N(z) g_{N1}(z)\tau_1(z)
\Bigr] \,  d z
 \: ,
\label{omegafin}
\end{eqnarray}
where $\Omega_0(T,\mu)$ contains all the terms independent of $\theta$
and the Tr$_1$ applies only to the layer 1, i.e., the first spacer 
layer.
If the system is invariant with respect to translations in the planes 
of atoms, or, if such a symmetry is restored by configuration averaging,
(\ref{omegafin}) can be written as
\begin{eqnarray}
&& \Omega(\theta,T,\mu) = 
\Omega_0(T,\mu)  
- \frac{1}{\pi}\:{\rm Im} \int_{-\infty}^{\infty}
\, f(E,T,\mu) \times \nonumber \\
& & \sum_{{\bf k}_{||}} \; {\rm tr \, ln}\: 
\Bigl[
1 - g_{1N}({\bf k}_{\|},z) \tau_N({\bf k}_{\|},z) 
    g_{N1}({\bf k}_{\|},z) \tau_1({\bf k}_{\|},z)
\Bigr] \,  d z 
 \: ,
\label{omegafink}
\end{eqnarray}
where tr means the trace over angular momentum indices $L=(\ell m)$ 
and the spin index $\sigma$.

\section{Useful mathematical tools}

Theoretical developments and many calculations are facilitated by the
partitioning technique and the two-potential formula applied to the 
Green function and its logarithm.

Let $P$ and $Q$ denote projection operators onto the complementary
subspaces (i.e. $P+Q = 1$). 
We denote the projections of matrices as 
$PAP = A_{PP}, PAQ = A_{PQ}$, etc., 
and $P/A$ means the inversion of $A_{PP}$ in the subspace
referring to projector $P$.
In most applications, $A=z-H$ or $A=P(z)-S$ and $G(z)=A^{-1}$.

The projections of the inverse $A^{-1}$ to the matrix $A$ are given by
\cite{Lowd}
\begin{eqnarray}
\label{invpart1}
(A^{-1})_{PP} &=& \frac{P}{A_{PP} - A_{PQ}\frac{Q}{A}A_{QP}} \, ,\\
\label{invpart2}
(A^{-1})_{QQ} &=& \frac{Q}{A_{QQ} - A_{QP}\frac{P}{A}A_{PQ}} \, ,\\ 
\label{invpart3}
(A^{-1})_{PQ} &=& - \frac{P}{A}A_{PQ}(A^{-1})_{QQ} = 
- (A^{-1})_{PP}A_{PQ} \frac{Q}{A} \, ,\\
\label{invpart4}
(A^{-1})_{QP} &=& - \frac{Q}{A}A_{QP}(A^{-1})_{PP} = 
-
 (A^{-1})_{QQ}A_{QP} \frac{P}{A} \, .
\end{eqnarray}

It is sometimes easier to invert the full matrix $A$ than its blocks.
In such a case the inverse partitioning is useful
\begin{eqnarray}
\label{backpart}
\frac{P}{A_{PP}} = (A^{-1})_{PP} \frac{P}{P-A_{PQ}(A^{-1})_{QP}}
= \frac{P}{P-(A^{-1})_{PQ} A_{QP}} (A^{-1})_{PP} \, .
\end{eqnarray}
This can be used to calculate the surface Green function of a
semi-infinite system from the Green function of the infinite 
system.

Partitioning technique also allows to simplify calculations
involving Tr ln of a matrix. 
The basic relation is
\begin{equation}
{\rm Tr \, ln}\, A = {\rm ln \, det}\, A \, .
\label{logdet}
\end{equation}
It then follows 
${\rm Tr \, ln}\, AB = {\rm Tr \, ln}\, A + {\rm Tr \, ln}\, B$, 
${\rm Tr \, ln}\, 1 = 0$, 
${\rm Tr \, ln} \, (A^{-1}) = - {\rm Tr \,ln}\, A$, and
${\rm Tr \, ln}\, [ (A - B)^{-1}] = - {\rm Tr \, ln}\, A - 
{\rm Tr \, ln}\, [ 1 - A^{-1}B ] $.
The Tr ln $A$ can then be partitioned as  
\begin{equation}
{\rm Tr \, ln} \, A = {\rm Tr}_{P} \, {\rm ln} \,[PAP] + 
{\rm Tr}_{Q} \, {\rm ln}\, [QAQ - QA\frac{P}{A}AQ].
\label{parttl}
\end{equation}
To prove~(\ref{parttl}), let us multiply the matrix $A$ 
by $L=1-A_{QP}(P/A)$ from left and by $R=1-(P/A)A_{PQ}$ from
right. 
The result is $LAR=A_{PP}+A_{QQ}-A_{QP}(P/A)A_{PQ}$.
Now using (\ref{logdet}), and the fact that det $[L]$ = det $[R]=1$
we find ({\ref{parttl}).
In a special, but important case, when $A_{PP}=P$ and 
$A_{QQ}=Q$ it holds
\begin{eqnarray}
{\rm Tr \, ln}\, A & = & {\rm Tr}_{P+Q} \, {\rm ln} \,
[P + Q + A_{PQ} + A_{QP}] \nonumber \\
& = & {\rm Tr}_{P} \, {\rm ln}\, [P - A_{PQ}A_{QP}] 
=  {\rm Tr}_{Q} \, {\rm ln}\, [Q - A_{QP}A_{PQ}] \, .
\label{redutl}
\end{eqnarray}

The Green function of a system described by the Hamiltonian
$H=H_0+v_1+v_2$, where $H_0$ is the unperturbed part, and 
$v_i (i=1,2)$ are perturbing potentials, is given by
$G=G^{(0)}+G^{(0)}TG^{(0)}$, where $G=(z-H)^{-1}$,
$G^{(0)}=(z-H_0)^{-1}$, and 
$T=V(1-G^{(0)}V)^{-1}$, where $V=v_1+v_2$.
The full T-matrix $T$ can be expressed in terms of the
t-matrices, $t_{i} = v_{i}(1-G^{(0)}v_{i})^{-1}, (i=1,2)$
and of the unperturbed resolvent $G^{(0)}$ by the
the two-potential formula 
\begin{eqnarray}
T=&& t_1\,[1-G^{(0)}t_2G^{(0)}t_1]^{-1}\,(1+G^{(0)}t_2)+
  t_2\,[1-G^{(0)}t_1G^{(0)}t_2]^{-1} \times \nonumber \\
& & (1+G^{(0)}t_1) \, .
\label{twopotgf}
\end{eqnarray}
It is derived in the following way.
Because
\begin{equation}
(1-A)[1 - (1-A)^{-1}AB(1-B)^{-1}](1-B) = 1-A-B \, ,
\end{equation}
it holds
\begin{eqnarray}
{\rm Tr \, ln}\, [1 - A - B] = {\rm Tr \, ln}\, [1 - A] + 
{\rm Tr \, ln}\, [1 - B] \nonumber \\
+ {\rm Tr \, ln}\, [1 - (1-A)^{-1}\,AB\,(1-B)^{-1}] \, .
\label{comptl}
\end{eqnarray}
By inserting $A=G^{(0)}v_{1}$ and $B=G^{(0)}v_{2}$ 
into (\ref{comptl}) one obtains (\ref{twopotgf}).
The two-potential formula for the Tr ln of the full
Green function
\begin{eqnarray}
{\rm Tr \, ln} \,G &=&
{\rm Tr \, ln} \,G^{(0)}[1-VG^{(0)}]^{-1} \nonumber \\
&=& {\rm Tr \, ln} \,G^{(0)} -
{\rm Tr \, ln} \,[1 - G^{(0)}v_{1} - G^{(0)}v_{2}] 
\nonumber \\
&=& 
{\rm Tr \, ln} \,G^{(0)} -
{\rm Tr \, ln} \,[1 - G^{(0)}v_{1}] - 
{\rm Tr \, ln} \,[1 - G^{(0)}v_{2}]
\nonumber \\
&& - {\rm Tr \, ln} \,[1 - G^{(0)}t_{1}G^{(0)}t_{2}] 
\label{twopotlog}
\end{eqnarray}
follows directly from (\ref{comptl}).

If the matrix $A$ is a function of a variable $z$ (complex in the 
general case), the derivative with respect to $z$ is given by
\begin{equation}
\frac{d}{dz} {\rm Tr \, ln}\, [A(z)] = 
{\rm Tr} \, \Bigl[\frac{d}{dz}A(z)\,A^{-1}(z)\Bigr] \, ,      
\label{deriva}
\end{equation}
provided that the matrix $A(z)$ is nonsingular.
This identity is used to derive the expression of the grandcanonical
potential $\Omega$ in terms of the auxiliary Green function (\ref{eq_gcp}) 
within the TB-LMTO.

The identity in (\ref{logdet}) is valid up to an integer multiple of
$2 \pi i$. 
Neglecting this fact can lead to serious errors. 
There is no panacea for this kind of difficulties, but in some
situations they can be avoided, for example by choosing the integration 
contour parallel to the imaginary axis, but this is not always possible.
In some cases the incremental procedure for calculating the ln det,  
$\ln f(z_{k+1}) = \ln f(z_k) + \ln \,[f(z_{k+1})/f(z_k)]$
in the spirit of an analytical continuation can be helpful, 
provided that the change of phase between two consecutive points $z_k$ 
is less than $2 \pi$.
To insure this, one has to choose a sufficiently small grid in $z$.

\section{Inversion of block-tridiagonal matrices}

We wish to compute $g = A^{-1}$ for a block-tridiagonal $A$.
The matrix $A$ is divided
into $N \times N$ square subblocks of the same dimension $m$,
from which non-zero are only $A_{k,k}$, $A_{k-1,k}$, and $A_{k,k-1}$.
The diagonal blocks are a sum of two terms: hermitean matrix and a
symmetric complex matrix.
They are always non-singular.
The off-diagonal blocks under the diagonal are equal to hermitean conjugate
of the corresponding blocks above the diagonal ($A_{k,k-1} = A^{+}_{k-1,k}$).
The methods based on repeated use of partitioning are particularly
efficient if only diagonal blocks, or four so-called 'corner'
blocks ($g_{1,1}, g_{N,N}, g_{1,N}, g_{N,1}$) are needed like
in the interlayer exchange coupling calculations.

First, four sequences of auxiliary matrices are calculated
\begin{eqnarray}
X_{N-k} &=& A_{N-k,N-k+1} (A_{N-k+1,N-k+1} - X_{N-k+1})^{-1}
A_{N-k+1,N-k}\, , \nonumber \\
& & \quad X_N = 0, \quad (k=1, \dots N-1) \, , \nonumber \\
Y_{k+1} &=& A_{k+1,k}(A_{k,k} - Y_k)^{-1} A_{k,k+1} \, ,
\quad Y_1 = 0, \quad (k=2, \dots N) \nonumber \\
Z_k &=& - (A_{k,k} - X_k)^{-1} A_{k,k-1}, \quad (k=2, \dots N) \nonumber \\
W_k &=& - (A_{k,k} - Y_k)^{-1} A_{k,k+1}, \quad (k=1, \dots N-1)  \, ,
\label{godf1}
\end{eqnarray}
that are used to compute the diagonal and off-diagonal blocks of $g$
\begin{eqnarray}
g_{k,k} &=& (A_{k,k} - X_k - Y_k)^{-1}  \, , \nonumber \\
g_{i,j} &=& Z_i \, g_{i-1,j} \quad \mbox{for} \quad i>j  \, , \nonumber \\
g_{i,j} &=& W_i \, g_{i+1,j} \quad \mbox{for} \quad i<j \, .
\label{godf2}
\end{eqnarray}
It can be proved that the numerical effort to evaluate the corner blocks
scales as $O(N m^{3})$.
The details, particularly the tests of efficiency can be
found in \cite{Godf}.


\begin{thebibliography}{99}

\bibitem{PBr} P.~Bruno, Phys. Rev. B {\bf 52}, 411 (1995).

\bibitem{JMa} J.~Mathon, M.~Villeret, A.~Umerski, R.B.~Muniz,
              J.~d'Albuquerque e Castro, and D.M.~Edwards, 
              Phys. Rev. B {\bf 56}, 11797 (1997).

\bibitem{IBM} See theoretical and experimental review articles in
              IBM J. Res. Development {\bf 42}, No. 1 (1998).

\bibitem{SHPK} M.~van Schilfgaarde, F.~Herman, S.S.P.~Parkin,
                and J.~Kudrnovsk\'y,  Phys. Rev. Lett. {\bf 74},
                4063 (1995).

\bibitem{ScH} M.~van Schilfgaarde and F.~Herman,
              Phys. Rev. Lett. {\bf 71}, 1923 (1993).

\bibitem{MSAJ}  S.~Mirbt, H.L.~Skriver, M.~Ald\'en, and B.~Johansson,
                Solid State Commun.~{\bf 88}, 331 (1993).

\bibitem{St} M.D.~Stiles, Phys. Rev. B {\bf 48}, 7238 (1993).

\bibitem{mft1} A.~Oswald, R.~Zeller, P.J.~Braspenning, and P.H.~Dederichs,
               J. Phys. F: Met. Phys. {\bf 15}, 193 (1985).

\bibitem{mft2} A.I.~Liechtenstein, M.I.~Katsnelson, V.P.~Antropov,
              and V.A.~Gubanov, J.~Magn.~Magn.~Mater.~{\bf 67}, 65 (1987).

\bibitem{ft} A.R.~Mackintosh and O.K.~Andersen, in {\it Electrons at the
             Fermi Surface}, Ch. 5.3., ed. M.~Springford (Cambridge
             University Press, Cambridge, England, 1980).

\bibitem{LlS} P.~Lloyd and P.V.~Smith, Adv. Phys. {\bf 21}, 69 (1972).

\bibitem{LNZD}  P.~Lang, L.~Nordstr\"om, R.~Zeller, and P.H.~Dederichs,
                Phys. Rev. Lett. {\bf 71}, 1927 (1993).

\bibitem{iecsgf}  J.~Kudrnovsk\'y, V.~Drchal, I.~Turek, and P.~Weinberger,
                  Phys. Rev. B~{\bf 50}, 16105 (1994).

\bibitem{iectq} V.~Drchal, J.~Kudrnovsk\'y, I.~Turek, and P.~Weinberger,
                Phys. Rev. B {\bf 53}, 15036 (1996).

\bibitem{iecvtx}  P.~Bruno, J.~Kudrnovsk\'y, V.~Drchal, and  I.~Turek,
                  Phys. Rev. Lett. {\bf 76}, 4254 (1996).

\bibitem{iectf} V.~Drchal, J.~Kudrnovsk\'y, P.~Bruno, and P.~Weinberger,
                Phil. Mag. B {\bf 78}, 571 (1998).

\bibitem{iectfa} V.~Drchal, J.~Kudrnovsk\'y, P.~Bruno, P.H.~Dederichs,
                 and P.~Weinberger, (to be submitted).

\bibitem{iecdb} J.~Kudrnovsk\'y, V.~Drchal, I.~Turek, M.~\v{S}ob,
                and P.~Weinberger, Phys. Rev. B~{\bf 53}, 5125 (1996).

\bibitem{iecas} J.~Kudrnovsk\'y, V.~Drchal, P.~Bruno, I.~Turek, 
                 and P.~Weinberger, Phys. Rev. B {\bf 54}, R3738 (1996).

\bibitem{iecmor} J.~Kudrnovsk\'y, V.~Drchal, C.~Blass, I.~Turek, 
                 and P.~Weinberger, Phys. Rev. Lett. {\bf 76}, 3834 (1996).

\bibitem{iecams} J.~Kudrnovsk\'y, V.~Drchal, R.~Coehoorn, M.~\v{S}ob, 
                 and P.~Weinberger, Phys. Rev. Lett. {\bf 78}, 358 (1997).

\bibitem{iecsor} P.~Bruno, J.~Kudrnovsk\'y, V.~Drchal, and I.~Turek, 
                 J. Magn. Magn. Mater. {\bf 165}, 128 (1997).

\bibitem{ieccap1} J.~Kudrnovsk\'y, V.~Drchal, P.~Bruno, I.~Turek, 
                  and P.~Weinberger, Phys. Rev. B {\bf 56}, 8919 (1997).

\bibitem{ieccap2} J.~Kudrnovsk\'y, V.~Drchal, P.~Bruno, R.~Coehoorn,
                  J.J. de Vries, K.~Wildberger, P.H.~Dederichs, and 
                  P.~Weinberger, MRS Symposium Proceedings, 
                  eds. J. Tolbin et al., Vol. {\bf 475}, 575 (1997).

\bibitem{AJ} O.K.~Andersen and O.~Jepsen, 
             Phys. Rev. Lett. {\bf 53}, 2571 (1984).

\bibitem{book} I.~Turek, V.~Drchal, J.~Kudrnovsk\'y, M.~\v{S}ob,
               and P.~Weinberger, {\it Electronic Structure of 
               Disordered Alloys, Surfaces and Interfaces}
               (Kluwer, Boston-London-Dordrecht, 1997)

\bibitem{tkd} See I.~Turek, J.~Kudrnovsk\'y, and V.~Drchal, these
              Proceedings.

\bibitem{agf} V.~Drchal, J.~Kudrnovsk\'y, L.~Udvardi, P.~Weinberger, 
              and A.~Pasturel, Phys. Rev. B {\bf 45}, 14328 (1992).

\bibitem{AHII} See, for example, A.~Messiah, {\it Quantum Mechanics},
               vol. II, (North-Holland, Amsterdam, 1969), Appendix C.
              
\bibitem{LNWZD} P.~Lang, L.~N\"ordstrom, K.~Wildberger, R.~Zeller, 
                and P.H.~Dederichs, Phys. Rev. B {\bf 53}, 9092 (1996).

\bibitem{FD} F.~Ducastelle, J. Phys. C: Solid State Phys. {\bf 8}, 
             3297 (1975).

\bibitem{SUWS} L.~Szunyogh, B.~\'Ujfalussy, P.~Weinberger, and
               C.~Sommers, Phys. Rev. B {\bf 54}, 6430 (1996).

\bibitem{WKu} B. Wenzien, J. Kudrnovsk\'y, V. Drchal, and M. \v{S}ob,
              J. Phys.: Condens. Matter~{\bf 1}, 9893 (1989).

\bibitem{sgf1} J.~Kudrnovsk\'y, I.~Turek, V.~Drchal, P.~Weinberger,
               N.E.~Christensen, and S.K.~Bose, Phys. Rev. B {\bf 46}, 
               4222 (1992).

\bibitem{sgf2} J.~Kudrnovsk\'y, I.~Turek, V.~Drchal, P.~Weinberger, 
               S.K.~Bose, and A.~Pasturel, Phys. Rev. B {\bf 47}, 
               16525 (1993).

\bibitem{sgf3} V.~Drchal, J.~Kudrnovsk\'y, and I.~Turek,
               Comp. Phys. Commun. {\bf 97}, 111 (1996).

\bibitem{pwsymm} P. Weinberger, Phil. Mag. B {\bf 75}, 509 (1997).

\bibitem{sform} J. Zabloudil, C. Uiberacker, U. Pustogowa, B. Blaas,
                L. Szunyogh, C. Sommers, and P. Weinberger, 
                Phys. Rev. B {\bf 57}, 7804 (1998).

\bibitem{sloncz} J.C.~Slonczewski, Phys. Rev. B {\bf 39}, 6995 (1989).

\bibitem{iecsc} D.M.~Edwards, A.M.~Robinson, and J.~Mathon,
                J. Mag. Mag. Mat. {\bf 140-144}, 517 (1995).

\bibitem{gmr} C. Blaas, P. Weinberger, L. Szunyogh, J. Kudrnovsk\'y,
              V. Drchal, P.M. Levy, and C. Sommers (submitted to
              J. Phys. I France). 

\bibitem{Godf} E.M.~Godfrin, J. Phys.: Condens. Matter {\bf 3}, 7843
               (1991).

\bibitem{WTS} P. Weinberger, I. Turek, and L. Szunyogh,
              Int. J. Quant. Chem. {\bf 63}, 165 (1997). 

\bibitem{iecsor1}  M.S.~Ferreira, J. Phys. Condens. Matter {\bf 9},
                   6665 (1997).

\bibitem{iecta} J.~d'Albuquerque e Castro, J.~Mathon, M.~Villeret,
                and A.~Umerski, Phys. Rev. B {\bf 53}, R13306 (1996).

\bibitem{Lan} C.~Lanczos, {\it Applied Analysis}, (Dover, New York,
              1988), p. 219.

\bibitem{PBrc}  P.~Bruno, J.~Magn.~Magn.~Mater. {\bf 164}, 27 (1996).

\bibitem{KKP} S.~Krompiewski, F.~S\"uss, and U.~Krey, 
              Europhys. Lett. {\bf 26}, 303 (1994).
              
\bibitem{NLZD} L.~N\"ordstrom, P.~Lang, R.~Zeller, and P.H.~Dederichs,
               Phys. Rev. B {\bf 50}, 13058 (1994).

\bibitem{Lowd} P.O.~L\"{o}wdin, J. Chem. Phys. {\bf 19}, 1396 (1951).

\end{thebibliography}
\end{document}